\documentclass[apj]{emulateapj}
\usepackage{apjfonts}
\usepackage{lscape}
\usepackage{graphicx, subfigure}

\defcitealias{lin07}{LM07}
\slugcomment{ApJ, 694, 992}  

\newcommand\xray{\hbox{X--ray} }
\newcommand\beq{\begin{equation}}
\newcommand\eeq{\end{equation}}

\newcommand{\totalgal}{139\ }
\newcommand{\totalcls}{110\ }

\shorttitle{SED of Cluster Radio Galaxies}
\shortauthors{Lin et al.}

\begin{document}

\title{Spectral Energy Distribution of Radio Sources in Nearby Clusters of Galaxies: 
Implications for Sunyaev-Zel'dovich Effect Surveys}

\author{
Yen-Ting Lin\altaffilmark{1,2},
Bruce Partridge\altaffilmark{3},
J.~C.~Pober\altaffilmark{3},
Khadija El Bouchefry\altaffilmark{4,3},
Sarah Burke\altaffilmark{5,3},\\
Jonathan N.~Klein\altaffilmark{3},
Joseph W.~Coish\altaffilmark{3},
and
Kevin M.~Huffenberger\altaffilmark{6}
}

\altaffiltext{1}{Princeton University Observatory, Princeton, NJ 08544; Departamento 
 de Astronom\'{i}a y Astrof\'{i}sica, Pontificia Universidad Cat\'{o}lica de Chile, Chile}
\altaffiltext{2}{Current address: Institute for the Physics and Mathematics of the Universe, University of Tokyo, Japan; yen-ting.lin@ipmu.jp
} 
\altaffiltext{3}{Department of Astronomy, Haverford College, Haverford, PA 19041; bpartrid@haverford.edu}
\altaffiltext{4}{Astrophysics and Cosmology Research Unit, 
  University of KwaZulu-Natal, South Africa}
\altaffiltext{5}{Swinburne University of Technology, Australia}
\altaffiltext{6}{Jet Propulsion Laboratory, California Institute of
 Technology, Pasadena, CA 91109;
Department of Physics, University of Miami, Coral Gables, FL 33146}

\begin{abstract}

To explore the high frequency radio spectra of galaxies in clusters,
we used NRAO's Very Large Array at four frequencies,
$4.9-43$ GHz, to observe \totalgal galaxies in low redshift ($z < 0.25$), X-ray detected, clusters. 
The clusters were selected from the survey conducted by Ledlow \& Owen, who provided 
redshifts and 1.4 GHz flux densities for all the radio sources. We find that more than half of the observed sources have steep microwave spectra as 
generally expected ($\alpha < -0.5$, in the convention $S\propto \nu^\alpha$). 
However, $60-70\%$ of the unresolved or barely resolved sources have 
flat or inverted spectra. Most of these show an upward turn in flux at $\nu > 22$ GHz, implying a higher flux than would be 
expected from an extrapolation of the lower frequency flux measurements. 
Our results quantify the need for careful source 
subtraction in increasingly sensitive measurements of the Sunyaev-Zel'dovich effect in clusters of galaxies (as currently being conducted by, for 
instance, the Atacama Cosmology Telescope and South Pole Telescope groups).
\end{abstract}


\section{Introduction}
\label{sec:intro}

The Sunyaev-Zel'dovich effect \citep[SZE;][]{sunyaev70} is a
powerful method for detecting clusters from observations of the cosmic microwave
background (CMB).  The hot electrons in the
intracluster medium inverse-Compton scatter the CMB photons,
distorting the CMB spectrum as seen in the direction of a cluster.
Because the SZE is redshift independent and is caused by the presence
of dense gas deep within the potential well of dark matter halos, SZE surveys can effectively 
detect high redshift clusters \citep[see e.g.,][for a recent review]{carlstrom02}
and are less confused by large scale structure than optical surveys.

Several microwave background experiments with mJy level sensitivity and 1--10 arcminute beams,
including the Atacama Cosmology Telescope (ACT\footnote{\url{www.physics.princeton.edu/act/index.html}}; \citealt{fowler07}),
the South Pole Telescope (SPT\footnote{\url{pole.uchicago.edu/}}; \citealt{staniszewski08,ruhl04}),
the Arcminute Microkelvin Imager (AMI\footnote{\url{www.mrao.cam.ac.uk/telescopes/ami/index.php}}; \citealt{zwart08,barker06}),
the Atacama Pathfinder Experiment SZ survey (APEX-SZ\footnote{\url{bolo.berkeley.edu/apexsz/}}),
and {\it Planck}\footnote{\url{www.rssd.esa.int/Planck/}}, will yield thousands of SZE clusters in the next few years;
in particular, 
all four ground experiments were already operational in 2007. The data from these surveys
will permit study of the mass function of clusters over cosmic epochs, 
a measurement which can elucidate the role of dark energy because structure growth
slows during dark energy domination.

As a probe of precision cosmology, a SZE survey has to control its
systematics, particularly regarding the correlation between the SZE signal
and cluster mass \citep[e.g., ][]{lin03a}. Radio point sources often found at or near cluster centers
pose serious challenges in this regard \citep{carlstrom02}. 
Powerful sources can overwhelm the cluster SZE signature \citep{cooray98,coble07},
and weaker, unresolved sources can collectively contaminate the SZE
signal \citep{pierpaoli04}. Clusters missed or affected this
way would distort the measurements of cosmological parameters from
SZE surveys, and it is therefore crucial to estimate the degree
of contamination due to radio sources.

Although at low frequencies (1.4--5 GHz) there have been extensive
studies of the radio galaxy population in clusters
(e.g., \citealt{ledlow96,miller01b,morrison03b,lin07}), it is not clear at
present how these sources behave at the frequencies ($\ge 15$
GHz) and flux levels ($\sim$ mJy) of on-going SZE surveys. Most of the forecasts
for future surveys therefore rely on large extrapolations either in
frequency or in flux level, and often both, from existing data 
(e.g., \citealt{toffolatti98,knox04,mwhite04,dezotti05}; however see \citealt{sadler08}
for recent observations at 95 GHz). 
For example, \citet[][hereafter \citetalias{lin07}]{lin07} use the observed spectral energy distribution
(SED) and spectral index distribution (SID) from 1.4 to 4.85 GHz to
estimate SZE survey contamination from the observed 1.4 GHz cluster
radio luminosity function. At 150 GHz, 
they estimate that
about 10\% of clusters of
mass $10^{14}-10^{15} M_\odot$ may host AGNs whose total fluxes exceed
that of the SZE signal. Although 
the AGN contribution can be detected and subtracted
by combining observations at different frequencies used in a
SZE survey, a more critical issue is to be able to quantify the
uncertainty about the fraction of clusters being lost or
contaminated at few percent level \citep{lima05}.
This extrapolation over two orders of magnitude in frequency (i.e., from 1.4 GHz to 150 GHz) is
highly uncertain, and points out the importance of understanding the actual
frequency dependence of these cluster sources.

An extensive follow
up of the 15 GHz 9C survey from 1.4 to 43 GHz \citep{bolton04}
clearly demonstrates that the SED of radio sources is highly
non-trivial, but does not focus on cluster radio sources. 
With sensitive observations toward 89 clusters over $0.1\le z <1$, 
\citet{coble07} determine the SID between 1.4 and 28.5 GHz. 
This is a major step toward understanding of the nature of the radio
sources in intermediate- to high-$z$ clusters. However, we note that their
sample is effectively selected against clusters
hosting bright radio sources. Furthermore, only a few of the radio sources are
spectroscopically confirmed cluster members.
Therefore it is not clear whether their sample is
representative of the cluster radio source population as a whole.

Here we present a systematic study of the spectral energy distribution of
{\it cluster} radio sources from 4.86 to 43.3 GHz, conducted with the Very Large
Array (VLA). \totalgal radio galaxies associated with \totalcls clusters at
$z<0.25$ are observed at three or four frequency bands {\it nearly simultaneously}, allowing better determination
of the spectral shape. Photometric data from the Sloan Digital Sky Survey (SDSS),
where available, are used to examine
correlations (if any) between the SID/SED and properties of the host galaxy and cluster,
such as color, luminosity and clustercentric distance.
Our survey improves upon previous studies in several aspects, including the
selection of cluster member galaxies based on available redshifts, the large
sample size, and
the near-simultaneous measurement of fluxes in all four bands.

The plan of the paper is as follows.
In \S\ref{sec:sample} we describe our cluster and radio galaxy sample.
The details of the observations and data reduction are provided in \S\ref{sec:obs}
and \S\ref{sec:redux}, respectively.
As the angular resolution of the observations at different frequencies is quite
different, we pay particular attention in comparing the fluxes in different bands;
the procedure is reported in \S\ref{sec:resolution}. We present the SED and SID of the
sources in \S\ref{sec:radio_results}, and the properties of the host galaxies and clusters
in \S\ref{sec:opt_results}. Based on these new results, we forecast the possible
contamination due to radio sources of SZE surveys in \S\ref{sec:sze}. We
conclude by summarizing our main findings and suggesting directions for
further work in \S\ref{sec:summary}.

Throughout this paper, we employ a flat $\Lambda$CDM 
cosmological model where $\Omega_M=1-\Omega_\Lambda=0.3$
and $H_0=70 h_{70}\,{\rm km\,s}^{-1} {\rm Mpc}^{-1}$.


\section{Cluster and Radio Galaxy Sample Selection}
\label{sec:sample}

Ledlow \& Owen conducted a 1.4 GHz survey of radio 
galaxies in $\sim 400$ clusters at $z< 0.25$ with a limiting sensitivity of 10 mJy,
and provided extensive redshift measurements for the host galaxies
\citep{ledlow95,ledlow96,owen95,owen97}.
Their cluster sample was drawn from the Abell catalogs \citep{abell58}, and was
restricted to area with reddening at $R$-band of less than 0.1 mag.
We further limited ourselves to \totalcls clusters that are detected in X-rays. The main
reasons for this requirement are: (1) as our ultimate goal is to make predictions for the radio source
contamination in SZE surveys, it is preferable to work with a cluster sample that is selected
in a similar fashion as in SZE surveys; and (2) the \xray emission provides a rough estimate of
the cluster mass, which is an important ingredient in our forecast for the SZE surveys.
As radio galaxies are rare, to maximize the sample size, we did not set any X-ray
flux limit as we compiled our cluster sample.
Based on Ledlow \& Owen's redshift catalog, \totalgal galaxies associated
with these clusters were selected as our  radio galaxy sample.

We note that Ledlow \& Owen surveyed the galaxies within 0.3 Abell radius of the cluster center
 (i.e., $\approx 0.64 h_{70}^{-1}$ Mpc), irrespective of the size (mass) of the clusters. Given the high
concentration in the spatial distribution of radio sources within clusters \citepalias{lin07},
their approach should include the majority of the sources associated with the clusters,
thus providing us with a representative initial sample of radio galaxies.
Using only sources projected within 40\% of the virial radius does not change the
derived SIDs or forecasts on the radio source contamination of the SZE (see \S\S\ref{sec:opt_results} \& \ref{sec:sze}).

In some cases we detect galaxies not in our initial sample
that we could confirm are cluster members on the basis of common redshift (see \S\ref{sec:bgn}).

\section{Observations}
\label{sec:obs}

Measurements in all four spectral bands were made at default VLA frequencies, centered at 43.3, 22.4, 8.5 and 4.9 GHz\footnote{Throughout the paper we refer to these frequency bands as Q, K, X, and C bands, respectively. Note that the 1.4 GHz channel is denoted as the L band.}. The observations discussed here were made in late October, 2005, with the VLA in a hybrid DnC configuration. In this configuration, the north-south baselines are on average longer than the east-west baselines, and as a consequence the synthesized beam is highly elliptical except for sources observed near the meridian at low elevation. During our runs, several antennas had been removed for repair or refitting; on average we had only 22 available, resulting in a $\sim 20\%$ 
reduction in sensitivity from the full array of 27. The first run, during the night Oct.~23-24, was carried out in mostly cloudy weather with poor atmospheric phase stability. We consequently elected to defer the 43 GHz observations to later runs. The high frequency Q-band observations were concentrated in a short run on Oct.~28 and a much longer run on Oct.~29-30 -- the latter in excellent, clear weather. The final short run on the night of 
Oct.~31 was used to obtain fill-in measurements on sources missed earlier at various frequencies.

\subsection{Calibration}

For all but the Oct.~31 run, our flux density scale was based on 1331+305 (3C286), for which NRAO specifies flux densities of 1.4554, 2.5192, 5.205 and 7.485 Jy at 43.3, 22.4, 8.5 and 4.9 GHz, respectively. 3C286 was not visible during our short run on Oct.~31; for these data we employed 3C48 as the primary flux calibrator, and carefully intercompared the flux densities obtained for sources and secondary calibrators observed in common on this day and earlier ones.
In the case of the two highest frequency bands, we employed standard software in the AIPS software package to import a model of the primary calibrators to take account of slight resolution effects in K and Q bands.

A variety of secondary (phase) calibrators were employed; we in general selected calibrator sources with reasonably flat spectra so that the same source could be used for observations in all four bands. Calibrators generally were within $\sim 15^\circ$ of all of our cluster sources.

Additional information on some instrumental parameters is provided in Table \ref{tab:para}. Note that the values for the synthesized beam shape are approximate, since the beam geometry depends on the declination and hour angle of the source.

\subsection{Fast Switching}
\label{sec:fs}

In the case of the 43 GHz observations, we employed fast switching between the source of interest and a nearby phase calibrator source. The integration times on source and calibrator were set to be approximately equal to or less than the atmospheric phase coherence time at 43 GHz. Rather than adjusting these integration times on the fly, we set them to be 100 sec on sources between calibrations, and 40 sec on nearby calibrators. For each galaxy observed, this cycle was repeated 3 times.

\begin{deluxetable*}{ccccc}

\tablecaption{Instrumental Parameters}
\tablewidth{0pt}

\tablehead{
\colhead{Frequency} & \colhead{Integration Time} &
\colhead{Typical Image Sensitivity} & \colhead{Image Pixel Size}  & \colhead{Synthesized Beam}\\
\colhead{(GHz)} & \colhead{(seconds)} &
\colhead{(mJy)} & \colhead{(arcsec)}  & \colhead{(approx.)}
}

\startdata

4.86 & 80 & 2.0 & $1.0$ & $8\arcsec\times 13\arcsec$\\
8.46 & 50 & 0.5 & $0.6$ & $4\arcsec\times 8\arcsec$\\
22.46 & 120 & 1.0 & $0.2$ & $2\arcsec\times 3\arcsec$\\
43.34 & $\sim 300$\tablenotemark{a} & 0.8 & $0.2$\tablenotemark{b} & $2\arcsec\times 3\arcsec$\tablenotemark{b}

\enddata

\tablenotetext{a}{fast switching employed (see \S\ref{sec:fs}).}
\tablenotetext{b}{for tapered images (see \S\ref{sec:taper}).}


\label{tab:para}

\end{deluxetable*}

\section{Data Reduction, Analysis, and Imaging}
\label{sec:redux}

The raw amplitude and phase data are flagged for shadowing of one antenna by another, interference, noisy correlators, weak antennas, and so on. In general, this flagging process removes only a few percent of the raw data. When data from the available antennas in the array are combined, the data are weighted by the inverse of the variance in the average signal.

Each source at each of the four frequencies is imaged using standard NRAO procedures in the AIPS software package. In forming the images, the pixel or cell size initially employed is $0.1\arcsec$, $0.2\arcsec$, $0.6\arcsec$ and $1.0\arcsec$ at 43.3, 22.5, 8.5, and 4.9 GHz, respectively. These values allow complete sampling of the synthesized beam even along its minor axis. In all cases, we make $1024^2$ images. The raw images are lightly cleaned of side lobes ($\sim 200$ iterations) again using standard NRAO procedures in AIPS. In most cases, when a source or sources are evident in the raw image, we clean first in a small area containing the source(s), then lightly clean the entire $1024^2$ pixel image. If no source is evident in the initial image, we simply clean lightly over the entire area. We have experimented with different levels of cleaning, and found no significant change in the flux densities of sources.

\subsection{Flux Density of Unresolved or Barely Resolved Sources}

Flux densities of evident, and unresolved or barely resolved, sources are determined by fitting a two-dimensional Gaussian to each image, using a standard process in AIPS (specifically, IMFIT). We report the integrated flux for each source. 
When no source is evident at or near the specified position, we compare $4\sigma$ ($\sigma$ is the local noise rms) to $2\sigma$ added onto the brightest flux per beam near the image center (within 50 pixels), and present the larger of the two as an upper limit.
%
%

\section{Effects of Resolution}
\label{sec:resolution}

As expected, many cluster radio sources show evident, resolved, structure at one or more of our observing frequencies. For resolved or irregular sources it is more difficult to obtain accurate fluxes; more importantly, the flux densities of resolved and complex sources are difficult to compare at different frequencies. For instance, the lobes of some of the classical FRII radio sources in our list are well delineated at 4.9 GHz, but only isolated hot spots in the lobes are visible at higher frequencies. In addition, the angular resolution of the VLA synthesized beam in the DnC configuration varies from $\sim 2\arcsec$ to $\sim 13\arcsec$ depending on frequency; much of the flux of extended sources is resolved out at higher frequencies. 
When flux is resolved out, only lower limits can be placed on spectral indices.

\subsection{Tapered 43 GHz Images}
\label{sec:taper}

Since we are most interested in the highest frequencies, 22 and 43 GHz, flux densities and the 22-43 GHz spectral index, we convolve our 43 GHz images with an elliptical Gaussian weighting profile to broaden the synthesized beam to match approximately the larger size of the synthesized beam at 22 GHz. This is done by applying a Gaussian weighting function to the $u-v$ data before imaging, again using a standard procedure in the AIPS task IMAGR. A $u-v$ taper of $45k\lambda$ and $135k\lambda$ provides a good overall match to the 22 GHz beam. For these tapered images we employ $0.2\arcsec$ cells, as for the 22 GHz images. By approximately matching the 22 and 43 GHz synthesized beams, we eliminate or reduce the problem of resolution and are able to compare fluxes from matched areas of the sky. Thus our 22-43 GHz spectral indices are unbiased values. 

Unless otherwise noted, all flux densities at 43 GHz are derived from these tapered images.

Because of the larger frequency ratio between 8.5 and 22 GHz, tapering the 22 GHz images to match the 8.5 GHz synthesized beam produces very noisy images (much of the $u-v$ data was strongly down-weighted), so we elect not to taper the 22 GHz images; see \S\ref{sec:sedofex} for further details. Hence spectral indices based on fluxes at 8.5 (or 4.9) GHz are lower limits, as noted above.

\subsection{Flux Density of Resolved or Irregular Sources}

In the case of irregularly shaped or clearly extended sources, we estimate the flux density within a rectangular region including all of the visible emission. These are figures cited in Table~\ref{tab:maindata} (see \S\ref{sec:radio_results}). Relatively few of the 22 and 43 GHz sources are complex enough to require this treatment.

\section{Observational Results}
\label{sec:radio_results}

We have observed \totalgal galaxies, and detected 136 in at least one band. The three that show no sign of a source at any of the frequencies are 0053$-$102B, 1108+410A, and 
1657+325B. These three are not included in our analysis. We note that the first and last of these three are weak 21 cm sources in Ledlow \& Owen's catalog. On the other hand, 1108+410A has a flux of 116 mJy in their catalog, but is very extended. In our 4.9 GHz image, we see faint traces of a source, but it is almost entirely resolved out even at $\sim 12\arcsec$ resolution.

For 111 galaxies we are able to measure flux in at least three bands. This subsample will allow
better determination of the spectral shape, and will be the focus 
of this section. 
Some of these galaxies have multiple components, and
in total we detect 140 radio sources associated with them.
Table~\ref{tab:break} records the detection statistics of our sources.
In Table~\ref{tab:maindata} we provide the available flux density measurements of all the sources.
Most blank entries in Table~\ref{tab:maindata} are for background sources (see \S\ref{sec:bgn}) far enough from the image centers so that they are not contained within the primary beams at the two higher frequencies we employed. In other cases, our runs on a particular source at a particular frequency were spoiled by weather or lost for other reasons. In a smaller number of cases, including for instance 0816+526,  sources seen independently at the higher frequencies were blended at 5 GHz, so that it was not possible to determine accurate flux densities at that frequency.

\begin{deluxetable}{ccccc}

\tablecaption{Detection Statistics}
\tablewidth{0pt}

\tablehead{
\colhead{} & \multicolumn{2}{c}{all morphologies} & \multicolumn{2}{c}{core/point-like} \\
\cline{2-3} \cline{4-5} \\
\colhead{detection\tablenotemark{a}} & \colhead{components} & \colhead{galaxies} & \colhead{components}  & \colhead{galaxies}
}

\startdata

all 4 & \phn87 & \phn75 & 57 & 57\\
$\ge 3$ & 140 & 111 & 73 & 73\\
$\ge 2$ & 185 & 133 & 83 & 83\\
$\ge 1$ & 192 & 136 & 86 & 85\\
\hline
total & 195\tablenotemark{b} & 139\tablenotemark{b} & 86 & 85


\enddata

\tablenotetext{a}{number of bands in which the sources are detected.}
\tablenotetext{b}{three sources detected at 1.4 GHz by \citet{owen97} were not detected at any frequency in our VLA observations.}

\label{tab:break}

\end{deluxetable}

\begin{figure*}[hbtp]
\begin{center}
\mbox{
	\leavevmode
	\subfigure
	{ 
	  \includegraphics[width=1.81in]{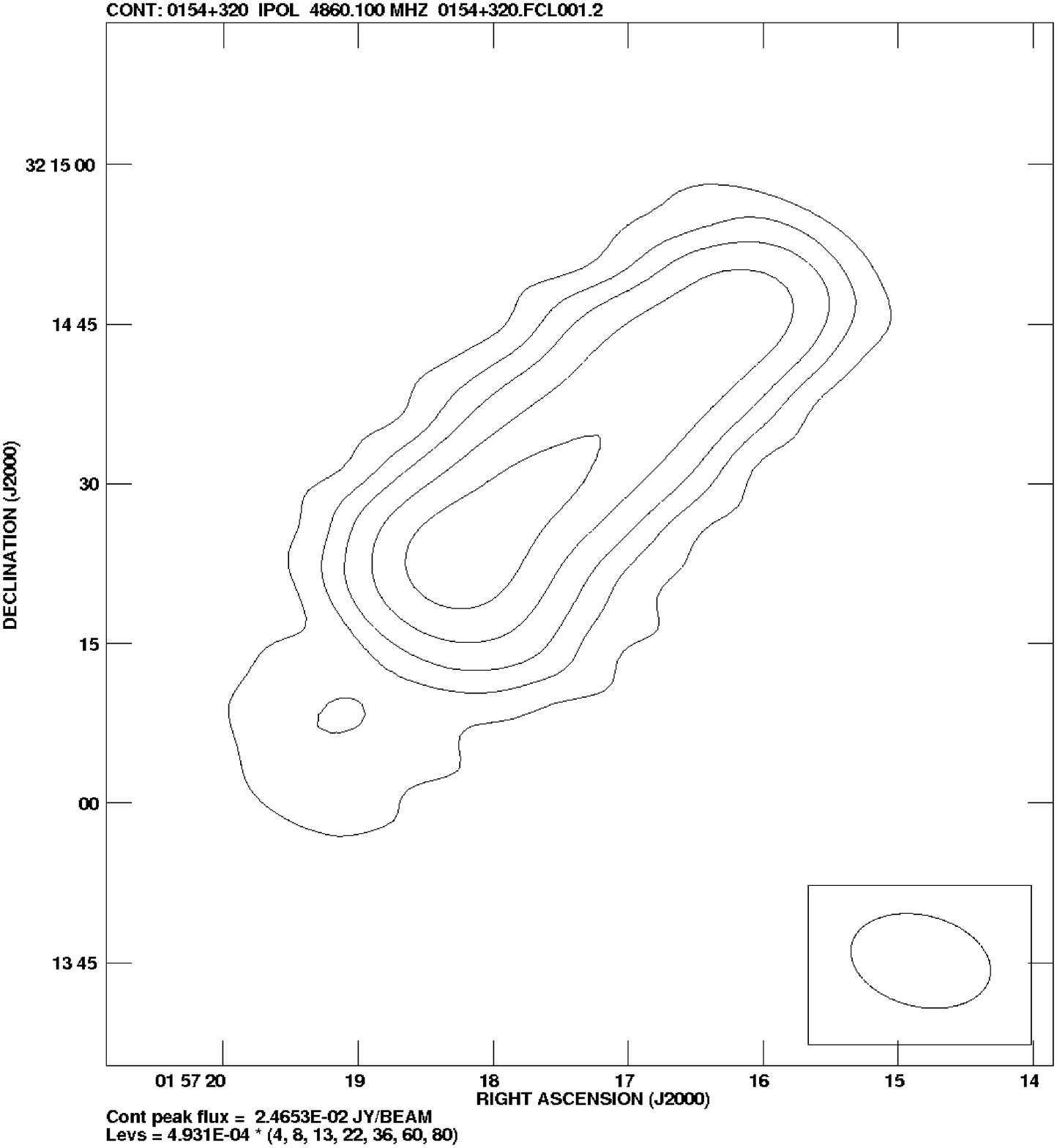} }

	\leavevmode
	\subfigure
	{ 
	  \includegraphics[width=2in]{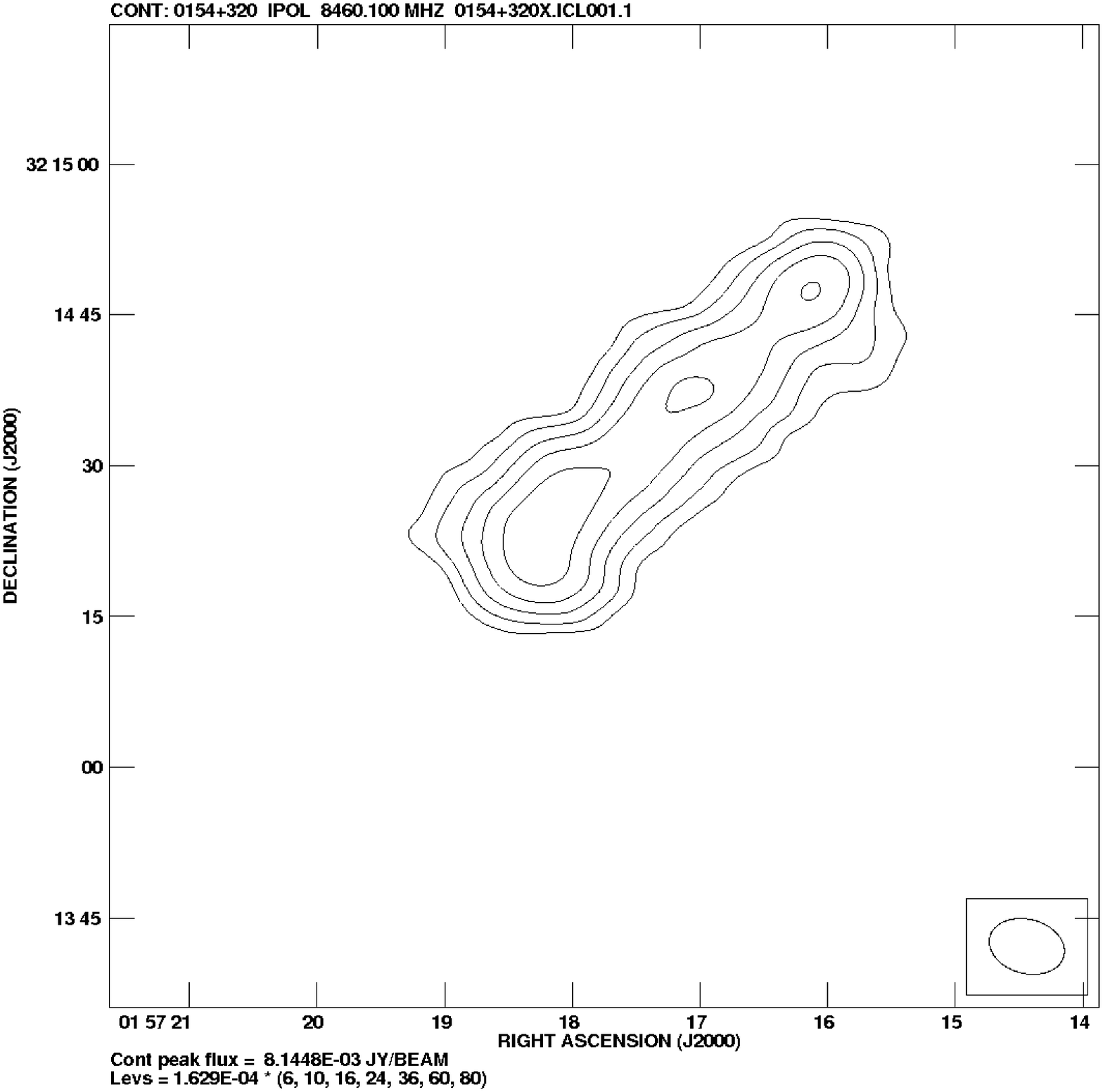} }

	\leavevmode
	\subfigure
	{ 
	  \includegraphics[width=1.82in]{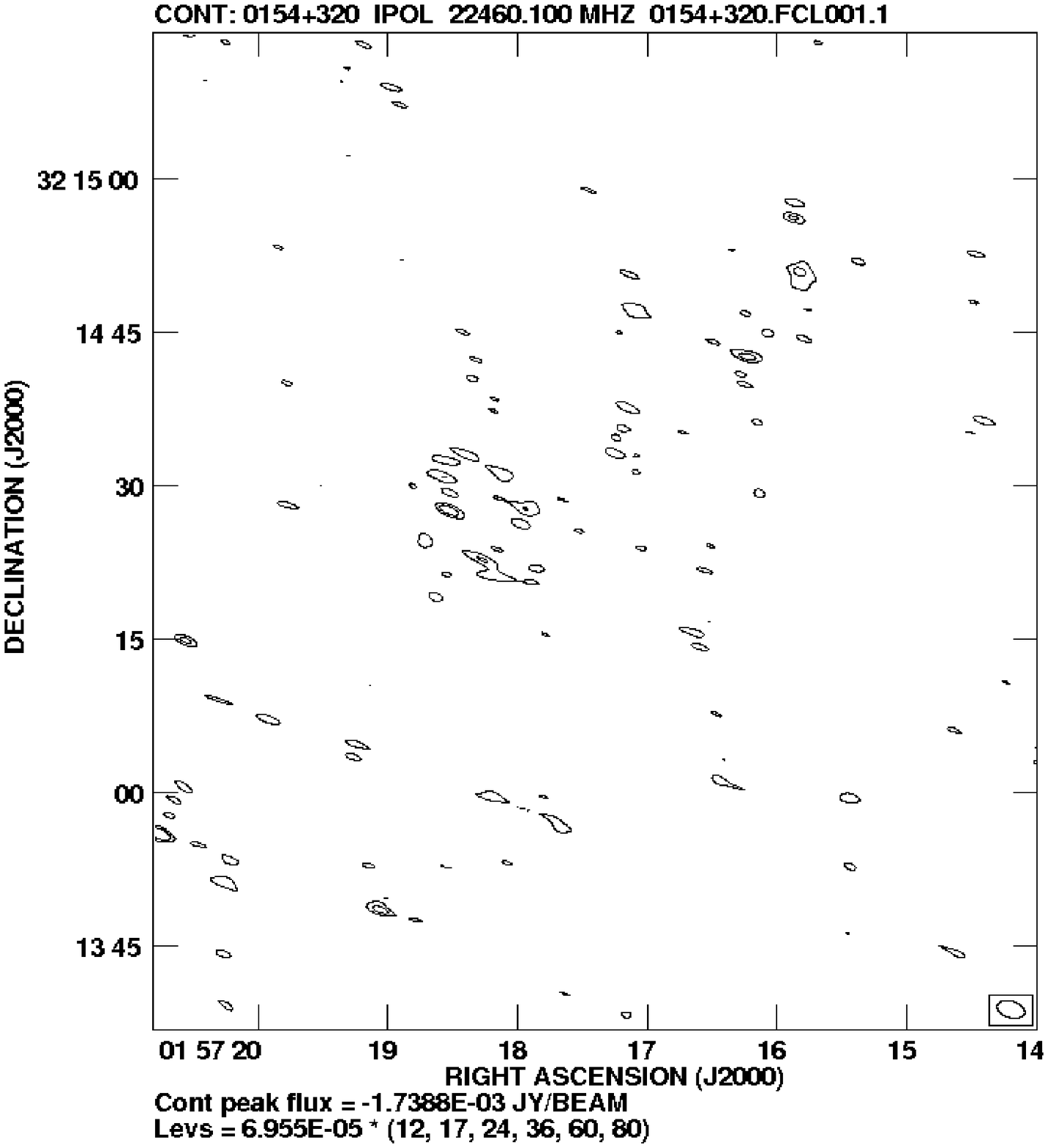} }
}	  
\end{center}
\caption{Left to right: a source (0154+320) imaged at C, X and K bands, showing the loss of flux due to resolution; beam profiles at each frequency are shown in the small box on the lower right of each panel.
Contours are selected to reveal the main source properties.
}
  \label{fig:cont}
\end{figure*}

\begin{figure}
\epsscale{0.85}
\hspace{-5mm}
\plotone{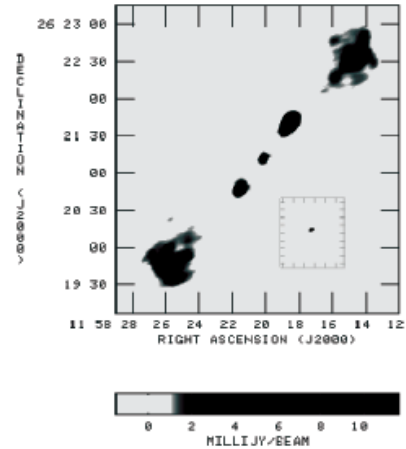}
\caption{ 
  Main figure:  the source $1155+266$ imaged at 8.5 GHz. Insert (to same scale): the same source at 22 GHz, showing no evidence for emission from the lobes; the core remains.
}
\label{fig:dd}
\end{figure}

\subsection{``Background'' Sources}
\label{sec:bgn}

In many of our images, especially those at low frequencies with their correspondingly larger solid angle, we by chance detect sources at a distance from the center of the field (or pointing position). Since our target galaxies are all at low redshift, these peripheral sources are presumably mostly background radio sources. 
We exclude from our analysis of the statistical properties of cluster radio galaxies all such ``background'' sources. 
Four of these ``background'' sources, however, have catalogued redshifts which show they are cluster members. These 4 are added to our sample of cluster galaxies in the subsequent analysis.

\subsection{Overall Properties of Cluster Radio Galaxies}

We now focus on the 111 cluster radio galaxies for which the spectral shape can be reliably traced,
since we have measurements at $\ge 3$ frequencies.
At the lowest frequency, 4.9 GHz, virtually all of the sources have complex structure. In $\sim 75\%$ of the sources, a clear core or small, barely resolved jet is visible. Even in these cases however, there is generally additional extended emission. In other sources, the cores or other resolved or barely resolved structures visible in higher frequency, higher resolution, images are unresolved or merged with diffuse structure in the 4.86 GHz images. An example is shown in Fig.~\ref{fig:cont}. This makes it difficult to isolate the cores at our lowest frequency, and to determine their flux densities for comparison with measurements at higher frequencies. In some cases, our best option is to compare the 4.9 GHz flux of an entire source with the sum of the flux densities of its components at the next highest frequency.

At the next highest frequency, 8.5 GHz, with a $4\arcsec \times 8\arcsec$ beam, cores and jets are more frequently resolved and isolated. On the other hand, we are resolving out some of the flux of extended features seen in the lower resolution 4.9 GHz image. As already noted, that means that the 8.5 GHz flux densities may be underestimated for large, extended sources and hence our calculated values of the 4.9-8.5 GHz spectral index are generally lower limits. On the other hand, the flux density determination for isolated, barely resolved cores and jets are more accurate and less influenced by background emission than is the case at 4.9 GHz.

At the two highest frequencies, because of the higher resolution, most of the extended structure seen at the two lower frequencies is resolved out and barely visible or not apparent (see Fig.~\ref{fig:dd}). As noted in \S\ref{sec:taper} above, we convolve the 43 GHz images to produce a synthesized beam matching that at 22 GHz. Thus we can fairly intercompare flux densities at the two higher frequencies, but it remains the case that spectral indices involving flux densities at either of the two lower frequencies will be lower limits.

\subsection{SEDs of Cores and Other Point-like or Barely Resolved Sources}

The very different resolution of our images at different frequencies makes it difficult to compare flux densities directly, and hence to determine SEDs, especially for complex or resolved sources. We therefore elect to concentrate on unresolved or barely resolved sources or the obvious cores within more complex structure. These sources are flagged in column 4 of Table~\ref{tab:maindata} with a ``C'' indicating a well defined core or ``P'' indicating an unresolved or barely resolved ``point-like'' source. We are not claiming that these sources are necessarily unresolved at $\sim 1\arcsec$ scale, but rather that they are sufficiently isolated and regular in appearance that accurate flux densities can be obtained. 

Of the 140 cluster sources for which the determination of a SED is possible,
73 or $52\%$
are either point-like or barely resolved in one or more of our three highest frequency images or have a clearly identifiable core at one or more of these same frequencies.

\subsection{SEDs of Extended Sources}
\label{sec:sedofex}

Because of the resolution effects discussed in \S\ref{sec:taper}, our SEDs and
spectral indices for extended sources are less certain. In general, as expected
\citep[e.g.,][]{deyoung02}, the lobes and diffuse structure show steep
spectral indices. In a few cases, as an experiment, we convolve the 8.5 GHz
images to match the resolution of the 4.9 GHz images to allow direct comparison
of fluxes.  The results are shown in Table~\ref{tab:taper}.  From the table, it is clear that
in these sample cases, at least, resolution is not significantly affecting the 8.5 GHz fluxes
in major ways, except for clearly resolved sources.


\begin{deluxetable*}{lclllr}

\tabletypesize{\scriptsize}
\tablecaption{Effect of Tapering on X-band Flux Measurements}

\tablehead{
\colhead{Source}	& \colhead{Extended}  &	\colhead{X Flux (un-tapered)}  &	\colhead{X Flux (tapered)}  &	\colhead{C Flux} &	\colhead{$\alpha_{CX}$ (tapered)}\\
\colhead{}	& \colhead{}  &	\colhead{(mJy)}  &	\colhead{(mJy)}  &	\colhead{(mJy)} &	\colhead{}
}

\startdata

0036$-$226B  &	   &	$59.850	\pm 0.430$&     \phn$61.770 \pm 0.520$ &	\phn$74.330 \pm 0.733$ &	$-0.334$ \\
0037+292 &	&	\phn$2.910 \pm 0.320$ &	\phn\phn$3.110 \pm 0.340$ &	\phn\phn$6.200 \pm 0.500$ &	$-1.245$ \\
0039$-$095B &	&	\phn$7.040 \pm 0.240$ &	\phn\phn$7.230 \pm 0.290$ &	\phn$13.260 \pm 0.644$ &	$-1.094$ \\
0100$-$221A &	&	\phn$4.495 \pm 0.520$ &	\phn\phn$3.180 \pm 0.260$ &	\phn\phn$6.110 \pm 0.570$ &	$-1.178$ \\
0119+193 (1) &	&	\phn$8.785 \pm 0.320$ &	\phn\phn$8.480 \pm 0.300$ &	\phn$11.880 \pm 0.610$ &	$-0.608$ \\
0124+189 &	Yes &	$75.440 \pm 1.910$ &	$132.900 \pm 2.830$ &	$310.000 \pm 4.260$ &	$-1.528$ \\
0139+073A &	&	\phn$4.898 \pm 0.230$ &	\phn\phn$4.710 \pm 0.320$ &	\phn\phn$6.270 \pm 1.040$ &	$-0.516$ \\
0909+161 &	&	$10.430 \pm 0.340$ &	\phn$11.380 \pm 0.410$ &	\phn$16.410 \pm 0.910$ &	$-0.660$ \\
1058+107 &	&	\phn$7.935 \pm 0.290$ &	\phn\phn$9.920 \pm 0.480$ &	\phn\phn$9.957 \pm 0.390$ &	$-0.007$ \\
1130+148 &	&	\phn$8.820 \pm 0.390$ &	\phn\phn$9.630 \pm 0.560$ &	\phn$13.680 \pm 0.410$ &	$-0.633$ \\
1132+492 &	&	$31.860 \pm 0.740$ &	\phn$34.640 \pm 0.900$ &	\phn$30.380 \pm 0.830$ &	$0.237$ \\
1201+282 &	&	\phn$1.870 \pm 0.430$ &	\phn\phn$1.680 \pm 0.550$ &	\phn\phn$2.930 \pm 0.310$ &	$-1.003$ \\
1301+195 &	&	$13.700 \pm 0.270$ &	\phn$12.790 \pm 0.290$ &	\phn$22.510 \pm 0.280$ &	$-1.020$ \\
1433+553 &	Yes &	$16.810 \pm 0.550$ &	\phn$21.870 \pm 0.710$ &	\phn$73.920 \pm 1.020$ &	$-2.197$ \\
1435+249 (1) & &	\phn$9.210 \pm 0.390$ &	\phn$10.160 \pm 0.500$ &	\phn$11.800 \pm 0.560$ &	$-0.270$ \\
2228$-$087 &	&	$10.480 \pm 0.312$ &	\phn$11.190 \pm 0.320$ &	\phn$10.893 \pm 0.670$ &	$0.049$ \\
2333+208 (1) &	&	\phn$6.601 \pm 0.270$ &	\phn\phn$7.680 \pm 0.290$ &	\phn$11.410 \pm 0.330$ &	$-0.714$ \\
2348+058 &	&	\phn$3.022 \pm 0.240$ &	\phn\phn$3.400 \pm 0.250$ &	\phn\phn$8.490 \pm 0.290$ &	$-1.651$

\enddata

\label{tab:taper}

\end{deluxetable*}

If only emission from lobes and extended structure were involved, the generally
steep spectra would ensure that most cluster radio sources would present
minimal problems for SZE measurements carried out at frequencies above, say, 90
GHz. 
The second-to-last column of Table~\ref{tab:maindata} lists estimated 90 GHz flux
densities based on the X band flux, assuming that the spectral index between 4.9 and 8.5 GHz 
can be directly extrapolated to 90 GHz. The last column of the Table is the estimate based
on the Q band flux where available, using the spectral index between 22 and 43 GHz.
Note the frequent substantial differences in extrapolated flux.

\subsection{SEDs of Cores}

However, we find that the cores and other unresolved or barely resolved
structures generally have flatter spectra, and in particular that many sources
exhibit a substantial change in spectral index at frequencies above 22 GHz.
This means that cluster radio sources may present a larger problem for
sensitive SZE measurements than might be expected from the extrapolation of low
frequency measurements (e.g., \citetalias{lin07}). For that reason, as well as
because of the difficulty of obtaining the fluxes of extended sources, we
concentrate on cores and other unresolved or barely resolved components.  We
will focus on the 73 sources that are detected in three or more bands with
these morphologies.  It is important to recall, however, that even for these
relatively compact and uncomplicated sources, resolution effects may cause us to miss some of the flux.  Since we use tapered Q band flux densities, $\alpha_{KQ}$ is nominally unaffected, but resolution may affect the spectral indices at lower frequencies.

\begin{figure}
\epsscale{1}
\plotone{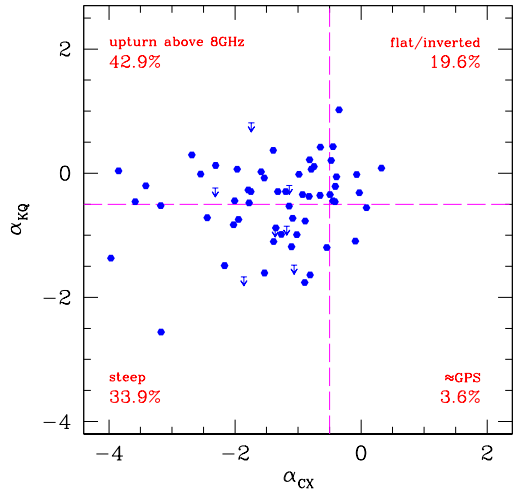}
\caption{ 
  The distribution of the spectral indices provides a way to quantify the relative proportions of different
  spectral shapes, as indicated in the four quadrants. We list the percentage of each type of spectral
  shape in the corresponding quadrants. Note that a large fraction of sources exhibit curvature in their
  spectra (e.g., the ``upturn'' type).
}
\label{fig:2color}
\end{figure}

An efficient way to quantify the distribution of the spectral shapes is through
the ``two-color'' plot \citep{sadler06}, shown in Fig.~\ref{fig:2color}, where
the spectral index between the C \& X bands ($\alpha_{CX}$) is plotted for 57 core/point-like sources against
that between the K \& Q bands ($\alpha_{KQ}$).  Throughout our analysis, we
adopt the notation for the spectral index such that a power-law spectrum is
described as $S_\nu \propto \nu^\alpha$.  Following the common practice of
referring to sources with $\alpha<-0.5$ as ``steep'', and ``flat/inverted''
otherwise, the $\alpha_{CX} - \alpha_{KQ}$ space is divided into four
quadrants, as delineated by the two dashed lines in Fig.~\ref{fig:2color}.
Starting from the first quadrant (upper-right corner) and going counterclockwise, the quadrants
contain sources with flat/inverted spectra, with spectra that 
turn flat above 8 GHz,
with steep spectra, and with spectra that peak around 10 or so
GHz [which we refer to as ``approximately gigahertz peaked spectrum'' ($\approx$ GPS) sources],
respectively.
The relative proportions of these types of spectral shape are shown in
Fig.~\ref{fig:2color}. 
We can also see that $\sim 84\%$ 
show a more positive high frequency spectral index
than low frequency spectral index, that is a flattening at frequencies above 8 GHz or so.
%
Only about one third of the sources have steep spectra from 4.9 to 43 GHz.  The
lack of correlations of the data points clearly suggests that
the spectral shape of the core/point-like sources is non-trivial.

If we include all 75 sources irrespective of their morphology, the relative
proportions of the four quadrants become 13.1\% (flat/inverted), 35.7\%
(upturn), 48.8\% (steep), and 2.4\% ($\approx$GPS).

With $\sim 100$ sources detected at 18 GHz with the Australian Telescope
Compact Array (ATCA), \citet{sadler06} study the distribution of the spectral
shapes with the two-color plot, where their low and high frequency indices are
based on 0.8 \& 5 GHz, and 8 \& 18 GHz fluxes, respectively.  We note that they
separate the flat/inverted sources from steep ones at $\alpha=0$; adopting the
same definition, we find that the great majority of our sources become steep
(73\%) and upturn (23\%).  This seems to suggest that our cluster sources
exhibit steeper spectra than theirs. However, the fact that the two samples
are selected at very different frequencies (1.4 v.s.~18 GHz) needs to be taken
into consideration. In addition, although the majority of their sources are
likely QSOs and BL Lac objects, most of them lack redshift information, which
makes it difficult to make a fair comparison (e.g., the nature of the sources and their
environments, as well as possible cosmological evolution).
Nevertheless, 
we agree with their conclusion that 
extrapolation of
fluxes to high frequencies (e.g., $\gtrsim 10$ GHz) based on low frequency
observations is not reliable.

\subsection{Spectral Index Distribution}

\begin{deluxetable}{ccc}

\tablecaption{Mean Spectral Indices}
\tablewidth{0pt}

\tablehead{
\colhead{Bands} & \colhead{All sources} &
\colhead{Cores/point-like}
}

\startdata

$5-8$ & $-1.64 \pm 0.10$ & $-1.31 \pm 0.10$\\
$8-22$ & $-1.20 \pm 0.07$ & $-0.88 \pm 0.09$\\
$22-43$ & $-0.98\pm 0.11$ & $-0.62 \pm 0.10$

\enddata

\label{tab:sid}

\end{deluxetable}

\begin{figure}
\epsscale{1}
\plotone{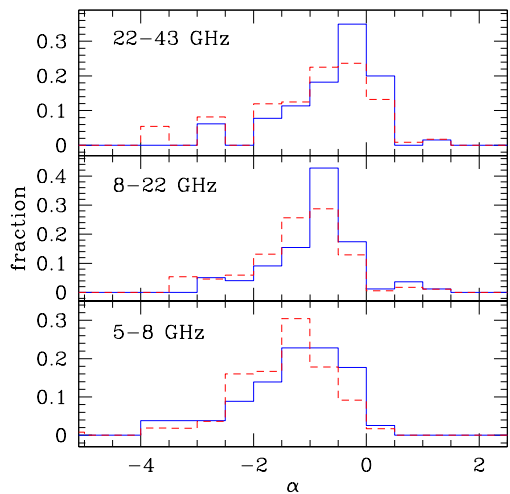}
\vskip -5mm
\caption{ 
  The spectral index distribution in three frequency bands. The solid histogram is the result
  when only sources with core/point-like morphology are used. The dashed histogram is obtained
  when all sources are included.
}
\label{fig:cx}
\end{figure}

Here we quantify the spectral index distribution in the $5-8$, $8-22$, and $22-43$
GHz bands.
An important aspect in estimating the SIDs is to deal with sources for which only an
upper limit in flux in one of the bands is available, which leads to upper or lower limits
of the spectral index. To accommodate such cases, we calculate the distribution with
the ASURV package \citep{feigelson85,isobe86}, which is based on 
survival statistics, 
a branch of statistics developed in actuarial estimates of human survival and mortality
(see e.g., \citealt{feigelson85} for a review).
The resulting SIDs are shown in Fig.~\ref{fig:cx}, where the solid histogram is
for sources with core/point-like morphology, and the dashed histogram is for
all sources.  We record the mean values of the indices in Table~\ref{tab:sid}.
As expected, the SIDs based on all the sources have a mean that is
more negative.

A recent study presents the spectral indices between 1.4 \& 28.5 GHz for
95 probable cluster radio sources \citep{coble07}. They find that the mean of
the index is $\sim -0.7$.  Because of the differences between the beam size of
the 1.4 GHz observations made by \citet{ledlow96} and ours, we do not attempt
to calculate an analog to their spectral index (e.g., $\alpha_{LK}$).
Furthermore, as the spectral shape tends to be complicated, it is not clear how
much predictive power an index spanning such a wide range in frequency would
have.

\section{Correlation of Spectral Indices and Properties of the Host Galaxies and Clusters}
\label{sec:opt_results}

Next we examine if there is any correlation between the spectral indices (from sources of
core/point-like morphology) and
the properties of the host galaxies or of the clusters.
In particular, we
consider the optical and radio luminosities, as well as the optical color, of
the host galaxies. As for the cluster-related properties, we look at the mass
and redshift of the clusters, and the projected radial distance to the cluster
center, which is determined from the emission peak of the intracluster gas.

\begin{figure}
\epsscale{1}
\plotone{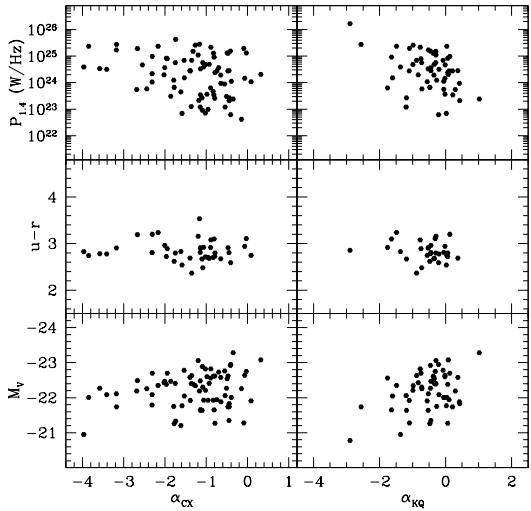}
\caption{ 
  Distribution of the spectral indices with respect to properties of host galaxies. We consider
  the absolute optical (V-band) magnitude, the optical color ($u-r$), and the radio power at 1.4 GHz
  of the hosts. No apparent correlation is found. 
}
\label{fig:gal}
\end{figure}

\begin{figure}
\epsscale{1}
\plotone{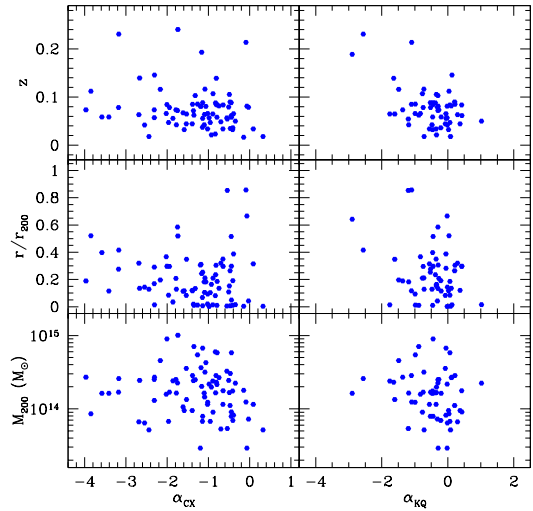}
\caption{ 
  Distribution of the spectral indices with respect to properties of host clusters. We examine the
  mass of the clusters, the distance to the cluster center (normalized by the virial radius of the cluster,
  $r_{200}$), and the redshift. As expected, radio sources concentrate towards cluster center. As
  in Fig.~\ref{fig:gal}, we do not find significant correlations.
}
\label{fig:cls}
\end{figure}

In Fig.~\ref{fig:gal} we show scatter plots between $\alpha_{CX}$/$\alpha_{KQ}$
and the galaxy properties. Fig.~\ref{fig:cls} is the corresponding plot for the
cluster properties.  A few points are worth commenting on both Figures.  First,
we note that the host galaxies are of moderate optical luminosity (recall that
$M_*=-20.8$ in the V-band), and are red in color ($u-r>2.2$).  Inspecting the
optical images of the host galaxies from SDSS confirms that most of the
galaxies are early type, of elliptical morphology.  The distribution of the 1.4
GHz luminosities ($P_{1.4}$) suggests that these galaxies are likely FRI-type radio-loud
AGNs.

The cluster mass $M_{200}$ is estimated from the X-ray luminosity
($L_X$)--virial mass relation \citep{reiprich02}. $M_{200}$ is defined as the
mass enclosed by $r_{200}$, a radius within which the mean overdensity is 200
times the critical density. Because of the scatter in the $L_X$--$M_{200}$
relation, our mass estimate is only accurate to $\lesssim 50\%$
\citep{reiprich02}.  Nevertheless, it is shown that $L_X$ is a unbiased mass
indicator \citep{reiprich06}.  As our main purpose is to find correlations with
the cluster mass, $L_X$ should suffice as a proxy for mass.  For each radio
source, we normalize its clustercentric distance by $r_{200}$, to account for
the difference in cluster mass. As Fig.~\ref{fig:cls} suggests, our clusters
span a range $>20$ in mass.  The majority of the sources are concentrated
toward the cluster center, which confirms several earlier findings (e.g.,
\citealt{morrison03b}; \citetalias{lin07}).

It is interesting to see that there appears to be no strong correlations between the
spectral indices and the host galaxies/clusters.  The Spearman's rank
correlation coefficients for all cases we examine are between $-0.1$ and $-0.34$,
indicating no significant correlations.  
The pair of properties that shows the strongest correlation
is that between $\alpha_{KQ}$ and $P_{1.4}$ (correlation coefficient $=-0.34$).
But this is mainly driven by a couple of sources that have the most negative spectral
index.

\citet{coble07} do not find any
difference between the spectral indices for sources in the inner and those in the outer
parts of the clusters, suggesting lack of correlation with clustercentric
distance, which is consistent with our finding here.
Considering the fact that spatial distribution of the low-power radio galaxies is
very concentrated towards cluster center (\citetalias{lin07}), this seems to suggest that
although being near the center of massive halos  increases the probability of accretion onto
the supermassive blackholes (e.g., high gas density or/and pressure from the
intracluster medium), the resulting emission is dominated by the small scale physics 
of the nucleus rather than by the cluster environment.

\section{Implications for SZE Surveys}
\label{sec:sze}

The main motivation to conduct the present study is to characterize the SED/SID
of radio sources associated with galaxy clusters, which can be used to assess
their effect on the detection and characterization of clusters through the SZE.
Simply put, the SIDs can be used to extrapolate the observed radio luminosity
function (RLF) at low frequencies to the frequency of an SZE experiment, which
in turn provides an estimate of the abundance of radio sources.

Our approach is similar to that of \citetalias{lin07}, and we refer the reader
to that paper for more details (\S7 therein). We will only provide an overview
of the method here. The basic idea is to use the (observed) RLF 
within clusters and groups to predict the number and flux of
radio sources expected in massive halos of given mass and redshift.
Specifically, the RLF gives the number density of radio sources which, when
multiplied by the volume of the halo, becomes the number of sources expected.
One can draw (Poisson) random numbers from it, and assign radio luminosities
according to the RLF. 
On the other hand, given the mass and redshift of a halo, one can predict its
SZE signal, which can be compared with the total fluxes from the radio sources.
By repeating this procedure for a large number of halos
of the same mass and redshift, one produces a radio galaxy catalog in a Monte
Carlo fashion, and can determine the fraction of clusters that are significantly
affected by the radio sources they host.
In \S\ref{sec:tech_details} we describe our scheme for extrapolating the RLFs,
and in \S\ref{sec:results} we present our estimates of the contamination of the SZE
due to cluster radio sources.

\subsection{Extrapolation of the Radio Luminosity Function}
\label{sec:tech_details}

\begin{figure}
\epsscale{1}
\plotone{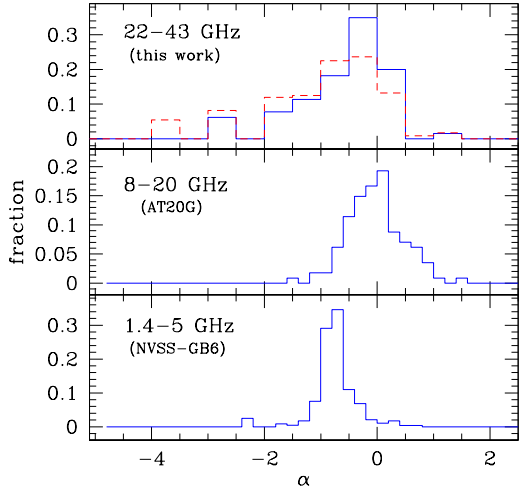}
\caption{ 
    The spectral index distribution in three frequency bands. {\it Top}: the SID in the $22-43$ GHz band
    as determined from our VLA data; this is identical to that shown in the top panel of Fig.~\ref{fig:cx}.
    {\it Middle}: the $8-20$ GHz SID determined from the AT20G survey, using data presented in \citet{sadler06}.
    The mean of the distribution is $\alpha_{XK,AT20G} = -0.028 \pm 0.046$.
    {\it Bottom}: the $1.4-5$ GHz SID based on data from the NVSS and GB6 surveys. 
    The mean is $\alpha_{LC,NVSS/GB6} = -0.754 \pm 0.024$. The details of the
    construction of SID($8-20$, AT20G) and SID($1.4-5$, NVSS/GB6) are described in the Appendix.
    Compared to the middle and bottom panels in Fig.~\ref{fig:cx}, the SIDs in this figure are more positive.
    In particular, as AT20G is a 20 GHz-selected survey, the resulting SID is biased towards flat-spectrum sources.
    }
\label{fig:sid2}
\end{figure}

The main ingredients in our method include: the 1.4 GHz RLF of radio sources residing
in massive halos, the SIDs between several frequencies, and a model
for the redshift evolution of the RLF. The 1.4 GHz RLF in units of space
density is measured in \citetalias{lin07}, and we give in \S\ref{sec:opt_results} the
distributions for $\alpha_{CX}$, $\alpha_{XK}$, and $\alpha_{KQ}$.
For SZE surveys operating at $\sim150$ GHz (e.g., ACT and SPT), for low redshift
sources, our modeling
requires a factor of $\sim 4$ extrapolation of the radio spectra in frequency
(i.e., from 43 GHz), which represents a dramatic improvement from that of
\citetalias{lin07}, which adapted an SID from 1.4 and 4.85 GHz measurements.
However, because of the mismatch between the angular resolutions in our images
at C, X, and K bands, the spectral indices we determine between these bands
may be lower limits, and thus the SIDs of $\alpha_{CX}$ and $\alpha_{XK}$ may
be biased towards negative values.
To assess the effects of the choices of SIDs, we will utilize other data sets
to determine the SIDs at $\nu \lesssim 20$ GHz.  To this end, we combine the
catalogs from the NVSS \citep{condon98} and GB6 \citep{gregory96} surveys to
measure the $1.4-4.85$ GHz SID, and use the results from the AT20G survey
\citep{sadler06} to measure the $8.5-20$ GHz SID.
We describe the construction of the matched NVSS/GB6 sample, as well as the
AT20G data, in the Appendix.
As the beam sizes of both NVSS and GB6 surveys are large ($45\arcsec$ and
$3.5\arcmin$, respectively), the flux, and in turn the spectral index
$\alpha_{LC}$ measurements,  should be reliable except for very extended
sources. On the other hand, the AT20G survey selects sources at 20 GHz, and
the resulting sample would be biased towards flat-spectrum sources. These SIDs
are shown in the middle and lower panels in Fig.~\ref{fig:sid2}.  The mean
values of $\alpha_{LC,NVSS/GB6}$ and $\alpha_{XK,AT20G}$ are $-0.754\pm 0.024$
and $-0.028 \pm 0.046$, respectively. 
Using the SID($1.4-4.85$, NVSS/GB6) and SID($8-20$, AT20G) rather than those
presented in \S\ref{sec:opt_results} will produce extrapolated RLFs with higher
amplitude (i.e., more radio sources), resulting in higher estimates of the contamination of SZE
signals.
We caution that the SIDs from NVSS/GB6 and AT20G are not limited to radio
sources {\it in} groups and clusters\footnote{We note that radio-loud AGNs are known to
reside in halos more massive than $\sim 10^{13} M_\odot$ or so (i.e., groups and cluster scale halos), based on their clustering properties \citep{mandelbaum08,wake08}.} (although the NVSS/GB6 sources are constrained to be
at $z<0.4$). However, incorporating these SIDs allows
us to explore the degree of AGN contamination of the SZE to a fuller extent.

\citetalias{lin07} measure the 1.4 GHz RLF for cluster radio sources.  We
transform that RLF to higher frequencies by convolving it with the spectral
index distribution via (\citetalias{lin07}) 
\begin{equation} 
\label{eq:tlf}
\phi_2\left( \log P_2 \right) = \int \phi_1 \left( \log P_2 +
\alpha\log(\nu_1/\nu_2) \right) f(\alpha_{12}) d\alpha_{12}, 
\end{equation}
where $\phi\equiv dn/d\log P$ is the RLF, and subscripts refer to two
frequencies 1 \& 2 ($\nu_2>\nu_1$).  The function $f(\alpha_{12})$ is the SID
between the two frequencies.
We have measured SIDs in several frequency bands: $1.4-5$, $5-8$, $8-20$
(or $8-22$), and $22-43$.  Depending on the frequency of the SZE experiment, we
may need to apply Eq.~\ref{eq:tlf} in several steps.  For example, 
the RLF at 5 GHz is obtained by extrapolating the 1.4 GHz RLF with $f(\alpha_{1.4,5})$, and
can be used in conjunction with $f(\alpha_{5,8})$ to obtain the 8 GHz RLF.
Convolving the latter with $f(\alpha_{8,20})$ [or $f(\alpha_{8,22})$] gives the RLF
at $10-30$ GHz range. Finally, 
RLFs at higher frequencies (e.g., 145 GHz)
are obtained by extrapolating the 22 GHz RLF with $f(\alpha_{22,43})$.

It is certainly preferable to utilize the full spectral shape from 5 to 43 GHz of our
sources for the extrapolation of the RLFs. We elect not to do so in the current analysis,
as our determination of spectral shape below 22 GHz may not be reliable. Instead, we
treat the spectral indices at different frequency bands as independent, and extrapolate the
RLFs in a piecewise fashion. This is justified given the lack of correlation of spectral indices
in the radio two-color diagram (Fig.~\ref{fig:2color}).

Ideally, one would extrapolate the RLF separately for the compact and extended components
of radio sources.
However, the 1.4 GHz RLF presented by \citetalias{lin07} is based on fluxes
from both the core and extended structures. 
Given that at low frequencies, the lobes usually dominate in flux over the cores
(e.g., Fig.~\ref{fig:dd}),
the core-only RLF
would have a smaller amplitude than the combined RLF. However, to determine the relative
proportion of the core-only and the lobe-only RLFs, one needs to carefully examine all radio
sources that contribute to the RLF, which is beyond the scope of the current analysis.
%

We note that the SID($1.4-4.85$) derived from the NVSS/GB6 surveys should be representative
for all sources with $\alpha_{1.4,4.85}\gtrsim -2$ (see Appendix), and therefore may result in an extrapolation of the 1.4 GHz RLF (to $\sim 5$ GHz) that appropriately takes into account the differences in the spectral shape of extended and compact sources. 
To further extrapolate to higher frequencies, 
we can use SIDs that are
known to be biased towards positive and negative values of spectral indices, thus giving the (presumably) full range of possible RLFs. Our forecasts on the radio source contamination of the SZE based on the RLFs will then reflect
the incomplete knowledge of the source spectral shapes.
To this end, for the four frequency ranges that we have determined the SIDs, we will employ a variety of SIDs to extrapolate the RLFs:
\begin{itemize}
\item $1.4-5$ GHz: SID($1.4-4.85$, NVSS/GB6).
\item $5-8$ GHz: SID($1.4-4.85$, NVSS/GB6), SID($5-8$, this work, point/core-like sources), or SID($5-8$, this work, all sources).
\item $8-22$ GHz: SID($8-22$, this work, all sources) or SID($8-20$, AT20G).
\item $22-43$ GHz: SID($22-43$, this work, point/core-like sources) or SID($22-43$, this work, all sources).
\end{itemize}

\subsection{Results}
\label{sec:results}

\begin{figure}
\epsscale{1.1}
\hspace{-2mm}
\plotone{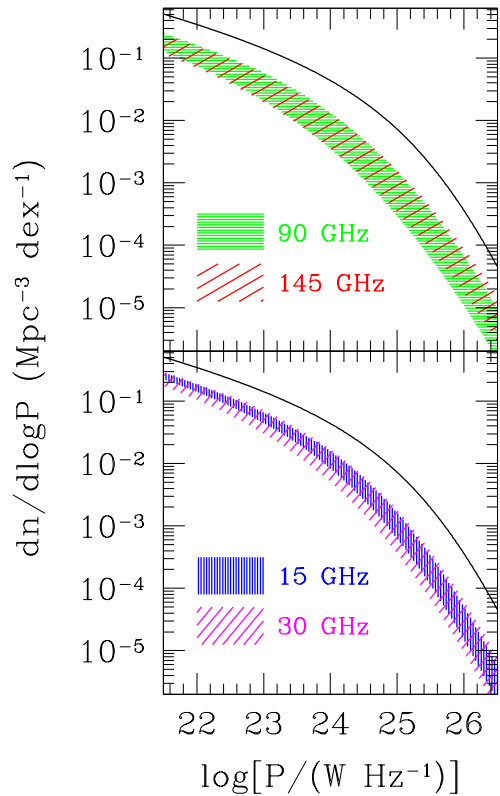}
\caption{ 
  Radio luminosity function (RLF) of radio-loud AGNs at four frequencies. 
  We transform the 1.4 GHz
  cluster AGN RLF from \citetalias{lin07} (measured within the virial radius; solid line) to 15 \& 30 GHz (bottom panel), and 90 \& 145 GHz (top panel), using Eq.~\ref{eq:tlf}.
  To account for uncertainties in the spectral shapes of the sources, we use a variety of SIDs (see \S\ref{sec:tech_details}) to extrapolate the RLF; the shaded regions enclose the probable range of the RLFs at these frequencies.
}
\label{fig:lfs}
\end{figure}

As explained in the previous subsection, using various combinations of SIDs we estimate the possible range of the RLFs given the uncertainties in the spectral shape of the radio sources. 
The $z\sim 0$ extrapolated RLFs (within $r_{200}$) at 4 frequencies are shown as the shaded regions in Fig.~\ref{fig:lfs}.  The bottom (top) panel shows the RLFs at 15 \& 30 GHz (90 \& 145 GHz).
At each frequency, the shaded region encloses the maximum and minimum of the RLFs resulted from the 12 SID combinations (6 for 15 GHz). For comparison, in both panels the solid curve is the 1.4 GHz RLF.

Compared to a similar plot presented in \citetalias{lin07} (Fig.~13 therein), a
dramatic change in the amplitudes of the extrapolated RLFs is seen.  At 145
GHz, at the luminous end, the upper envelop of the RLF is about a
factor of $60$ less than that estimated in \citetalias{lin07}.  This is due to
the combined effect of (1) the use of several SIDs in different frequency bands
in the present analysis (so that the results are not strongly dependent on one
single SID), and (2) that the SID used in \citetalias{lin07} may be biased to
positive indices, as non-detections at 4.85 GHz during the matching of sources
between 1.4 and 4.85 GHz were not properly taken into account (see Appendix for
more discussion).

\begin{figure}
\epsscale{1}
\plotone{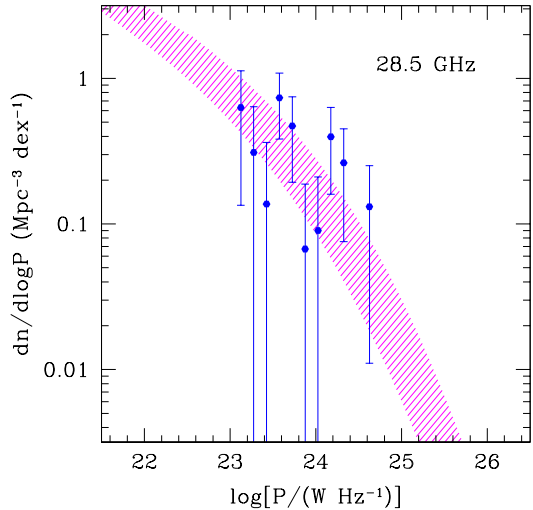}
\caption{ 
  Comparison of our extrapolated RLF (shaded region) at 28.5 GHz with observations (points), for sources within $r_{2000}$ (about $1/3$ of the virial radius, $r_{200}$) in 37 clusters at $0.1\le z\le 0.3$. The data is taken from \citet{coble07}. 
  Note that the extrapolated RLFs shown in Fig.~\ref{fig:lfs} are for regions within $r_{200}$; for comparison with observations restricted to $r_{2000} \approx r_{200}/3$, we have scaled the 28.5 GHz RLF within $r_{200}$ by a constant factor that takes into account the different spatial distribution of mass and radio sources [assuming the two components follow the \citet{navarro97} profile with concentration of 5 and 30, respectively]. 
  }
\label{fig:30ghz}
\end{figure}

To check our extrapolation scheme, we show in Fig.~\ref{fig:30ghz} a comparison between our extrapolated and the observed RLFs at 28.5 GHz, using data from \citet{coble07}.
We restrict ourselves to sources in the 37 clusters in the \citet{coble07} sample which have redshifts in the range $0.1-0.3$, and for which X-ray observations are available, in order to match the selection criteria for our sources. Next, as the \citet{coble07} observations are made at the BIMA and OVRO arrays, we need to take account of the primary beam size of these arrays ($6.6\arcmin$ and $4.2\arcmin$, respectively). We therefore include only the 27 radio sources that fall inside a projected radius of $r_{2000} \approx 0.33 r_{200}$, as this radius roughly matches the FWHM beam of BIMA for clusters in the redshift range $0.1-0.3$.
%
Using the source counts at 30 GHz from \citet{knox04}, we estimate that roughly 11 sources could be
background objects. Given this uncertainty and the small number of sources, the RLF is not well determined. 
Furthermore, it is difficult to evaluate the impact of \citeauthor{coble07}'s cluster selection on the resulting RLF (if anything, the amplitude of the RLF should be higher, because their cluster sample is selected {\it against} those hosting bright point sources).
Nevertheless, it is reassuring to find that
there is a general agreement between the data points and our extrapolation (shaded region).

We note that the redshift evolution of radio sources in clusters is an
unresolved issue.  Our sample is limited to $z<0.25$, and currently there is no
consensus as how radio galaxies evolve in massive halos (see \citetalias{lin07}
for discussion). We have acquired C, X, and K band data for a sample of radio
galaxies in $\sim 10$ intermediate-redshift clusters. Better constraints on the
redshift evolution based on these new data will be presented in a future
publication. 
In the current analysis, we will assume a pure density evolution of the form $\phi(z) \propto \phi(z=0)
(1+z)^\gamma$, with $\gamma=1$, which corresponds to a factor of 2 increase of the density at $z\approx 1$.
Such an evolution is derived from an analysis of the cluster radio source evolution from the
Red sequence Cluster Survey (Roscioli \& Gladders 2008, in preparation; M.~Gladders, 2008,
private communication), and is much milder than what is assumed in LM07. The contamination
of the SZE due to radio sources based on the present analysis is therefore much smaller when compared to the forecast presented in LM07.

\begin{figure}
\epsscale{1.1}
\plotone{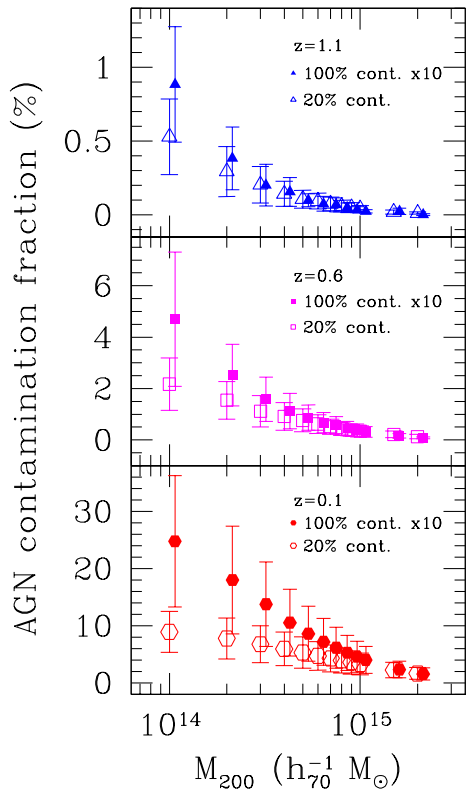}
\caption{ 
  Fraction of clusters which host enough radio-loud AGNs such that their SZE signal measurements at 145 GHz
  may be contaminated. We consider cases where the fluxes of the AGNs are at least a fraction $q$
  of the SZE signal, $S_{AGN} \ge q |S_{SZE}|$ (for the case where SZE signal is a temperature 
  decrement), with $q=0.2$ (open symbols) and $q=1$ (solid symbols). Note the $q=1$ contamination fraction is {\bf multiplied by a factor of 10} for clarity of presentation.
  From bottom to top we examine clusters at $z=0.1$, 0.6, and 1.1. A mild density evolution of the radio sources is assumed (corresponding to a factor 2 increase at $z=1$ compared to $z=0$). The contamination fraction is quite small (always $<10\%$), and decreases with both cluster mass and redshift. 
The errorbars show the $1\sigma$ range of possible degrees of contamination, reflecting our incomplete knowledge of the spectral shape of the radio sources.
}
\label{fig:sz}
\end{figure}

Now, given the mass and redshift of a dark matter halo, we can estimate the degree of
contamination as follows.  For a halo, we denote the total fluxes from radio
sources as $S_{AGN}$, and the SZE signal as $S_{SZE}$.  Using our Monte Carlo
scheme to generate a large number of radio sources in massive
halos, the fraction of halos for which
$S_{AGN}$ is a significant fraction $q$ of $|S_{SZE}|$ can be calculated. We
consider two cases, $q=0.2$ and $q=1$, corresponding to 20\% and 100\%
contamination.  We show in Fig.~\ref{fig:sz} the resulting AGN contamination
fraction (ACF) at 145 GHz, which is the proportion of the clusters expected to host radio
galaxies whose flux is $S_{AGN}\ge q |S_{SZE}|$, as a function of cluster mass.
The 3 panels show the results at $z=0.1$, 0.6, and 1.1 (bottom to top). In each panel,
the open points refer to the case of $q=0.2$, while the solid points show the $q=1$ ACF {\bf multiplied by
a factor of 10} (for better presentation). 
Using different combinations of SIDs, we have constructed 12 145 GHz RLFs, resulting in a range of degree of contamination (at a given halo mass and redshift). 
While the points show the mean value of the contamination fractions, the error bars indicate the $1\sigma$ range based on the 12 estimates.

The general trend shown in Fig.~\ref{fig:sz} is that ACF decreases as cluster mass and redshift increase.
For $q=0.2$ contamination, the most affected clusters are those nearby, at $\sim 10\%$ level for $M_{200}=10^{14} M_\odot$ clusters, reaching to $1-2\%$ for $M_{200}\ge 10^{15} M_\odot$ ones.
At $z=0.6$, an epoch close to the peak of cluster redshift distribution, the $q=0.2$ ACF is reduced to $<2\%$ at $10^{14} M_\odot$, and becomes negligible towards high mass end. At $z=1.1$ the $q=0.2$ ACF is always at sub-percent level.
Finally, the proportion of clusters that are affected by AGNs to 100\% is a factor of
$4-5$ smaller than the above estimates.
For completeness, we note that 
about $0.4-4$\% ($1-7$\%) of clusters in the mass range $10^{14}-10^{15} M_\odot$ 
at $z\sim 0.6$ may be contaminated to 10\% (5\%) level (i.e., $q=0.1$ and 0.05,
respectively). At $z=1.1$, these values become $0.04-1\%$ ($0.2-2\%$).

We have provided a framework for estimating the abundance of radio-loud AGNs in
halos. To better determine the impact of radio sources in SZE surveys, however, it is
necessary to carry out mock observations that take into account the properties of the
telescope and receiver system (e.g., angular resolution, sensitivity, frequency; see \citealt{sehgal07}), 
as well as the
auxiliary observations (e.g., availability of multiwavelength data).

\section{Summary and Future Work}
\label{sec:summary}

We have presented a study of the spectral energy distribution of radio sources
in a large sample of nearby clusters ($z<0.25$).  For 139 sources selected at
1.4 GHz and spectroscopically confirmed to be members of the clusters, we use the
VLA to measure the flux densities at 4.9, 8.5, 22, and 43 GHz (C, X, K, and Q bands)
nearly simultaneously, and determine the distribution of the SED.  Sources
with extended morphology may be resolved out at high frequencies (i.e.,
reduction in flux due to the higher angular resolution of interferometer),
making the determination of the spectral shape nontrivial.
We have downgraded the resolution of our 43 GHz images to match the resolution
at 22 GHz, thus enabling reliable 
comparisons of fluxes at these two frequencies
(\S\ref{sec:taper}); it is more difficult to match the resolution between the
other frequency intervals, and therefore our measurements of the spectral
indices involving the two lower frequencies (e.g., between 8.5 and 22 GHz,
$\alpha_{XK}$, where $S\propto \nu^\alpha$) are {\it lower} limits.
The flux measurement of point-like (or barely resolved) sources, or ``cores''
embedded in extended sources, on the other hand, is more straightforward.

Our main findings are the following:\\
1. For $\sim 70$ core/point-like sources that are detected in at least three
frequencies, we study the distribution of the spectral shape via the
``two-color'' diagram (Fig.~\ref{fig:2color}), and find that the spectral
shape cannot be described by simple power-laws for the bulk of the sources.
About 60\% of sources have $\alpha_{KQ}>\alpha_{CX}$, indicating a flattening
of the spectral shape above 8 GHz or so; only $1/3$ of the sources have steep
spectra in the entire range from 4.9 to 43 GHz.

\noindent 
2. We determine the spectral index distribution using survival
statistics that take non-detections (upper limits) into account. The
results are shown in Fig.~\ref{fig:cx} and Table~\ref{tab:sid}.  The compact
sources are found to have ``flatter'' spectral shape than the extended sources.

\noindent
3. The spectral indices do not correlate with properties of host galaxies or
clusters, such as the color and luminosity of the galaxies, the radio
luminosity at 1.4 GHz, the distance of the host galaxy to the cluster center,
and the mass of the host clusters. This result agrees with previous studies,
and suggests that the radio emission may be dominated by the small scale
physics of the nucleus, rather than by the cluster environment.

\noindent
4. In an attempt to estimate the contamination of the SZE signal due to radio
point sources in cluster surveys, we make use of the spectral index
distributions in several frequency bands to extrapolate the well-measured RLF
at 1.4 GHz to the frequencies employed by several on-going radio/millimeter
wave experiments.  As the extrapolation depends on the SIDs employed, we
bracket the possible range of the predicted RLFs by using SIDs in intermediate
frequency bands (e.g., $8-20$ GHz) that are known to be biased 
in opposite ways.
The amplitude of the resulting RLFs at $\nu\ge 30$ GHz is
in general $5-20$ times lower compared to that at 1.4 GHz.  Under the
assumption that the RLF follows a pure density evolution with redshift of the
form $\phi(z) \propto \phi(z=0) (1+z)^\gamma$, such that the abundance of
sources at $z=1$ is twice the local value, we find that the fraction of
clusters that may be seriously affected by point sources is quite small; at the
cluster mass scales close to the detection limits of the on-going surveys
(e.g., $2-3\times 10^{14} M_\odot$), and at the redshift where we expect the
experiments to detect most of the clusters (i.e., $z\sim 0.6$), $\lesssim 2\%$
of the clusters will be contaminated to $20\%$ level or above (that is, the total
fluxes from AGNs are at least 20\% of the SZE signal).

There are two aspects that need to be improved for a better forecast
within our analysis framework.
Currently,
the largest uncertainty in our modeling 
is the redshift evolution of cluster radio galaxies.  If not properly accounted
for, any unexpected evolution of the radio sources may be misinterpreted as
changes in the cluster mass function, and cause errors in the determination of
the properties of the dark energy.  With our on-going VLA survey of cluster
radio galaxies at intermediate redshift ($0.3\le z\le 0.8$), we plan to address
this issue in a future publication.

In addition,
in our forecast, it is implicitly assumed that the number of radio galaxies ($N_{RG}$) 
a cluster can host is proportional to the cluster mass ($M_{200}$). If, instead,
$N_{RG} \propto M_{200}^s$ with $s<1$, we would overestimate the AGN contribution
in high mass clusters. To check this assumption, one needs to determine the halo
occupation distribution for radio galaxies. 
To this end, we have attempted to construct the halo occupation distribution
of radio galaxies, using a large sample of radio galaxies in the local Universe
(Lin et al.~2008, in preparation).

Mainly because of the very mild redshift evolution of the radio sources we adopt (which is based
on the results from the RCS survey; Roscioli \& Gladders 2008, in preparation), we find that
radio sources do not cause a substantial degree of contamination to the SZE signal. To control the systematics in the on-going and future SZE cluster surveys, it is thus
crucial to understand the contamination due to the dusty IR sources. 
Rapid progress has been made on this regard \citep[e.g.,][]{righi08,fernandez08}. It would
be important to perform an assessment of contamination due to both radio and IR sources
within a single framework, which is the goal of our research in the near term.

\acknowledgments 

We thank David Spergel, Jill Knapp,  Anna Sajina, Joe Mohr, Tom Crawford, and Heinz Andernach
for helpful comments and suggestions.
We are grateful to Mike Gladders for sharing results on the evolution of radio sources in clusters
prior to publication, and to an anonymous referee for constructive comments that have improved
the paper. YTL thanks IH for constant encouragement and support.
YTL acknowledges supports from the Princeton-Cat\'{o}lica Fellowship, 
NSF PIRE grant OISE-0530095, FONDAP-Andes,
and the World Premier International Research Center Initiative, MEXT, Japan.
BP acknowledges support from NSF grant AST-0606975.  
A portion of this work was performed at JPL, under a contract with NASA.

The National Radio Astronomy Observatory is a facility of the National Science
Foundation operated under cooperative agreement by Associated Universities,
Inc.

Funding for the Sloan Digital Sky Survey (SDSS) and SDSS-II has been provided
by the Alfred P. Sloan Foundation, the Participating Institutions, the National
Science Foundation, the U.S. Department of Energy, the National Aeronautics and
Space Administration, the Japanese Monbukagakusho, and the Max Planck Society,
and the Higher Education Funding Council for England. The SDSS Web site is
http://www.sdss.org/.

The SDSS is managed by the Astrophysical Research Consortium for the
Participating Institutions. The Participating Institutions are the American
Museum of Natural History, Astrophysical Institute Potsdam, University of
Basel, University of Cambridge, Case Western Reserve University, The University
of Chicago, Drexel University, Fermilab, the Institute for Advanced Study, the
Japan Participation Group, The Johns Hopkins University, the Joint Institute
for Nuclear Astrophysics, the Kavli Institute for Particle Astrophysics and
Cosmology, the Korean Scientist Group, the Chinese Academy of Sciences, Los Alamos National Laboratory, the Max-Planck-Institute for
Astronomy, the Max-Planck-Institute for Astrophysics, New Mexico
State University, Ohio State University, University of Pittsburgh, University
of Portsmouth, Princeton University, the United States Naval Observatory, and
the University of Washington.

This research has made use of the NED and BAX databases, and the data products
from the NVSS and GB6 surveys.

\appendix

\section{Spectral Index Distributions from NVSS/GB6 and AT20G Surveys}

Here we describe the construction of the spectral index distributions,
SID($1.4-5$, NVSS/GB6) and SID($8-20$, AT20G), shown in Fig.~\ref{fig:sid2}.

The NRAO VLA Sky Survey \citep[][NVSS]{condon98} is a 1.4 GHz survey covering the sky
north of $\delta=-40^\circ$, with a resolution of $45\arcsec$. The nominal
detection limit is 2.5 mJy.  The Green Bank 4.85 GHz (GB6) survey
\citep{gregory96} used NRAO's (former) 91m telescope to survey the sky within
$0^\circ<\delta<75^\circ$, with a resolution of $3.5\arcmin$, and a detection
threshold of 18 mJy.  We first match the NVSS source catalog to the the
spectroscopic sample of SDSS DR6 (with a conservative matching radius of
$10\arcsec$), and limit the combined sample to $z<0.4$, as we are interested in
the radio galaxies in the local Universe.  We then cross correlate the
NVSS/SDSS sample with GB6 (again using a conservative matching radius of
$1\arcmin$), keeping all unmatched NVSS/SDSS sources (for which we can derive upper limits on $\alpha_{LC}$).  Because of the differences
in the angular resolution of the two radio surveys, we further limit ourselves
to NVSS sources for which there are no neighboring sources from NVSS within a
radius of $4\arcmin$.  This is to ensure that both surveys measure the
``total'' flux from the sources,
and to avoid sources that might be blended in the lower resolution GB6 survey.
Finally, to account for the differences in the
detection limits, we set a high flux cut (100 mJy) for NVSS sources so that we
can be sure to include all sources with $\alpha_{LC}>-1.4$. Of the resulting
292 NVSS sources, 9 are not detected in GB6, and we assign 18 mJy as the upper
limit in the 4.85 GHz flux for these sources.
The SID from this sample is shown in the lower panel in Fig.~\ref{fig:sid2}.
We note our result is not sensitive to the flux cut applied to the NVSS
sources, or on the requirement for the ``isolatedness'' of the sources.
Neither setting the cut to 200 mJy (so that we are complete for sources with
$\alpha_{LC}>-1.9$) nor including sources with neighbors closer than $4\arcmin$
changes the mean value of the SID beyond the one sigma level.

The Australia Telescope Compact Array is conducting a large survey at 20 GHz
(AT20G) that will eventually cover the sky south of $\delta=0^\circ$. A bright
source catalog based on observations up to 2004 is reported by
\citet{sadler06}. We use the 114 sources stronger than 100 mJy at 20 GHz that
are also detected at 8.6 GHz to construct the SID, and show the result in the
middle panel of Fig.~\ref{fig:sid2}.  Because the sources are selected at 20
GHz, the SID is biased towards positive values. Furthermore, the redshifts for
the majority of the sources are not available, and thus the result may not be
representative of the restframe $8-20$ GHz SID. Nevertheless, this sample
provides a distribution that is at the opposite extreme compared to that from
our VLA observations, and therefore the two SIDs should bracket the true
distribution.

\bibliographystyle{apj}

\clearpage
\LongTables 
\begin{landscape} 

\begin{deluxetable*}{lrrcrrrrrrcccrr}
\tighten

\tabletypesize{\tiny}
\tablecaption{Cluster Radio Sources}
\tablewidth{0pt}

\tablehead{
\colhead{Name\tablenotemark{a}} & \colhead{RA} & \colhead{Dec} & \colhead{Type\tablenotemark{b}}  &
\colhead{$z$\tablenotemark{c}} & \colhead{$f_L$\tablenotemark{d}} &
\colhead{$f_C$} & \colhead{$f_X$} & \colhead{$f_K$} & \colhead{$f_Q$} &
\colhead{$\alpha_{CX}$} & \colhead{$\alpha_{XK}$} &
\colhead{$\alpha_{KQ}$} & \colhead{$f_{90,CX}$\tablenotemark{e}} & \colhead{$f_{90,KQ}$\tablenotemark{f}}\\
\colhead{} & \colhead{(J2000)} & \colhead{(J2000)} & \colhead{}  &
\colhead{} & \colhead{(mJy)} &
\colhead{(mJy)} & \colhead{(mJy)} & \colhead{(mJy)} & \colhead{(mJy)} &
\colhead{} & \colhead{} & \colhead{} & \colhead{(mJy)} & \colhead{(mJy)}
}

\startdata

0036$-$226B &      9.7842 & $   -22.3338$ & C & 0.0654 & 129 &  $74.33 \pm 0.73$ &   $59.85 \pm 0.43$ &   $39.12 \pm 0.95$ &   $37.68 \pm 1.16$ &   $ -0.39$ &  $ -0.44$ &  $ -0.06$ &  $23.77$ & $36.12$\\
0037+209 &      9.9397 & $    21.2256$ & $-1$ & 0.0622 & 143 &  $10.43 \pm 2.05$ &   $<0.66$ & $<1.77$ & $<1.49$ & $< -4.97$ &  \nodata & \nodata & \nodata & \nodata\\
0037+292 &     10.1180 & $    29.5561$ & P/C & 0.0716 & 12 &  $6.20 \pm 0.50$ &   $2.91 \pm 0.32$ &   $2.55 \pm 0.95$ &   $<1.43$ & $ -1.36$ &  $ -0.14$ &  $< -0.88$ &  $ 0.12$ & \nodata\\
0039+211 &     10.4230 & $    21.4026$ & C & 0.1030 & 670 &  $101.70 \pm 2.20$ &   $48.84 \pm 0.55$ &   $24.70 \pm 0.65$ &   $20.34 \pm 0.80$ &   $ -1.32$ &  $ -0.70$ &  $ -0.30$ &  $ 2.15$ & $16.35$\\
0039$-$095B &     10.4603 & $    -9.3031$ & C & 0.0556 & 55 &  $13.26 \pm 0.64$ &   $7.04 \pm 0.24$ &   $2.11 \pm 0.48$ &   $<1.85$ & $ -1.14$ &  $ -1.24$ &  $< -0.20$ &  $ 0.47$ & \nodata\\
0039$-$095A &     10.4509 & $    -9.2840$ & $-1$ & 0.0556 & 48 &  $20.98 \pm 2.09$ &   $5.20 \pm 0.90$ &   $<2.57$ & $<1.48$ & $ -2.51$ &  $< -0.72$ &  \nodata & $ 0.01$ & \nodata\\
0039$-$097 &     10.4592 & $    -9.4296$ & C & 0.0556 & 82 &  \nodata & $6.86 \pm 0.32$ &   $4.10 \pm 0.48$ &   $3.95 \pm 0.79$ &   \nodata & $ -0.53$ &  $ -0.06$ &  \nodata & $ 3.79$\\
0043+201(1) &     11.6233 & $    20.4671$ & C & 0.1053 & 563 &  $19.63 \pm 1.56$ &   $14.93 \pm 0.50$ &   $<1.68$ & $2.20 \pm 0.76$ &   $ -0.49$ &  $< -2.24$ &  \nodata & $ 4.65$ & \nodata\\
0043+201(2) &     11.6222 & $    20.4681$ & C & 0.1053 & 563 &  $19.63 \pm 1.56$ &   $14.93 \pm 0.50$ &   $10.63 \pm 0.48$ &   $8.48 \pm 0.65$ &   $ -0.49$ &  $ -0.35$ &  $ -0.35$ &  $ 4.65$ & $ 6.58$\\
0046+011 &     12.1665 & $     1.4302$ & C & 0.0632 & 68 &  $25.57 \pm 0.88$ &   $5.76 \pm 0.46$ &   $2.63 \pm 0.57$ &   $3.19 \pm 0.98$ &   $ -2.68$ &  $ -0.80$ &  $  0.29$ &  $ 0.01$ & $ 3.96$\\
0047+241 &     12.4245 & $    24.4451$ & C & 0.0818 & 200 &  $25.23 \pm 1.08$ &   $16.68 \pm 0.41$ &   $7.01 \pm 0.53$ &   $7.52 \pm 1.07$ &   $ -0.75$ &  $ -0.89$ &  $  0.11$ &  $ 2.86$ & $ 8.14$\\
0047+242A(1) &     12.4365 & $    24.5003$ & $-1$ & 0.0818 & 24 &  $11.86 \pm 1.29$ &   $5.62 \pm 0.57$ &   $1.03 \pm 0.45$ &   $<2.12$ & $ -1.35$ &  $ -1.74$ &  $<  1.10$ &  $ 0.23$ & \nodata\\
0047+242A(2) &     12.4348 & $    24.5008$ & C & 0.0818 & 24 &  $11.86 \pm 1.29$ &   $4.82 \pm 0.49$ &   $1.89 \pm 0.54$ &   $<2.12$ & $ -1.62$ &  $ -0.96$ &  $<  0.18$ &  $ 0.20$ & \nodata\\
0050$-$220(1) &     13.3629 & $   -21.7503$ & $-1$ & 0.0587 & 97 &  $11.80 \pm 1.36$ &   $6.78 \pm 0.55$ &   \nodata & \nodata & $ -1.00$ &  \nodata & \nodata & $ 0.64$ & \nodata\\
0050$-$220(2) &     13.3567 & $   -21.7366$ & $-1$ & 0.0587 & 97 &  $25.57 \pm 0.89$ &   $13.94 \pm 0.51$ &   $6.73 \pm 0.71$ &   $3.21 \pm 0.76$ &   $ -1.09$ &  $ -0.75$ &  $ -1.13$ &  $ 1.05$ & $ 1.40$\\
0053+261A(1) &     13.9624 & $    26.4065$ & $-1$ & 0.1971 & 1327 &  $82.20 \pm 1.79$ &   $21.74 \pm 0.68$ &   $<2.40$ & $<2.21$ & $ -2.40$ &  $< -2.26$ &  \nodata & $ 0.08$ & \nodata\\
0053+261A(2) &     13.9580 & $    26.4131$ & $-1$ & 0.1971 & 1327 &  $81.37 \pm 2.06$ &   $24.13 \pm 0.95$ &   $<2.06$ & $<2.21$ & $ -2.19$ &  $< -2.52$ &  \nodata & $ 0.14$ & \nodata\\
0053$-$102B & 13.9668  & $-9.9847$ & $-1$ & 0.0534 & 28 &  $<1.95$ & $<0.45$ & $<2.40$ & $<1.85$ & \nodata & \nodata & \nodata & \nodata & \nodata\\
0053$-$015 &     14.1068 & $    -1.2623$ & C & 0.0444 & 1764 &  $104.00 \pm 4.30$ &   $44.33 \pm 1.16$ &   $19.94 \pm 1.15$ &   $18.98 \pm 0.75$ &   $ -1.54$ &  $ -0.82$ &  $ -0.08$ &  $ 1.17$ & $17.96$\\
0053$-$016 &     14.0074 & $    -1.3430$ & $-1$ & 0.0444 & 1095 &  $282.00 \pm 8.00$ &   $106.53 \pm -1.00$ &   $7.77 \pm 1.74$ &   $<1.62$ & $ -1.75$ &  $ -2.69$ &  $< -2.40$ &  $ 1.68$ & \nodata\\
0100$-$221A &     15.6880 & $   -21.9041$ & P & 0.0566 & 138 &  $6.11 \pm 0.57$ &   $4.50 \pm 0.52$ &   $<3.22$ & \nodata & $ -0.55$ &  $< -0.34$ &  \nodata & $ 1.22$ & \nodata\\
0108+173 &     17.7649 & $    17.6521$ & P & 0.0638 & 12 &  $6.44 \pm 0.61$ &   $3.51 \pm 0.28$ &   $<1.96$ & $<1.90$ & $ -1.09$ &  $< -0.60$ &  \nodata & $ 0.26$ & \nodata\\
0110+152 &     18.2483 & $    15.4913$ & C & 0.0444 & 719 &  $31.50 \pm 0.60$ &   $14.53 \pm 2.84$ &   $4.97 \pm 0.65$ &   $6.33 \pm 0.60$ &   $ -1.39$ &  $ -1.10$ &  $  0.37$ &  $ 0.54$ & $ 8.30$\\
0119+193(1) &     20.5867 & $    19.5881$ & P & 0.0544 & 20 &  $11.88 \pm 0.61$ &   $8.79 \pm 0.32$ &   $5.25 \pm 0.46$ &   $2.40 \pm 0.54$ &   $ -0.54$ &  $ -0.53$ &  $ -1.20$ &  $ 2.43$ & $ 0.99$\\
0122+084 &     21.2819 & $     8.6994$ & P & 0.0498 & 51 &  $4.03 \pm 0.61$ &   $3.32 \pm 0.32$ &   $1.24 \pm 0.57$ &   $2.42 \pm 0.47$ &   $ -0.35$ &  $ -1.01$ &  $  1.02$ &  $ 1.45$ & $ 5.13$\\
0123$-$016A &     21.4345 & $    -1.3795$ & C & 0.0180 & 910 &  $77.60 \pm 11.50$ &   $20.02 \pm 1.67$ &   $10.93 \pm -1.00$ &   $6.84 \pm 0.57$ &   $ -2.44$ &  $ -0.62$ &  $ -0.72$ &  $ 0.06$ & $ 4.04$\\
0123$-$016B &     21.5027 & $    -1.3451$ & P & 0.0180 & 3270 &  $107.53 \pm 9.16$ &   $128.70 \pm 3.61$ &   $121.70 \pm 1.40$ &   $128.38 \pm 1.00$ &   $  0.32$ &  $ -0.06$ &  $  0.08$ &  $276.79$ & $136.33$\\
0124+189 &     21.7266 & $    19.2145$ & C & 0.0420 & 1345 &  $310.00 \pm 4.26$ &   $75.44 \pm 1.91$ &   $24.98 \pm 1.23$ &   $24.75 \pm 0.96$ &   $ -2.55$ &  $ -1.13$ &  $ -0.01$ &  $ 0.18$ & $24.49$\\
0139+073A &     25.4977 & $     7.6806$ & P & 0.0616 & 43 &  $6.27 \pm 1.04$ &   $4.90 \pm 0.23$ &   $2.07 \pm 0.64$ &   $2.74 \pm 0.81$ &   $ -0.44$ &  $ -0.88$ &  $  0.43$ &  $ 1.71$ & $ 3.76$\\
0139+073B(2) &     25.5183 & $     7.6506$ & $-1$ & 0.0616 & 27 &  $9.67 \pm 1.84$ &   $3.28 \pm 0.38$ &   $2.23 \pm 0.84$ &   $<1.51$ & $ -1.95$ &  $ -0.40$ &  $< -0.60$ &  $ 0.03$ & \nodata\\
0149+359(1) &     28.1932 & $    36.1518$ & $-1$ & 0.0163 & 81 &  $22.42 \pm 1.09$ &   $10.78 \pm 0.47$ &   $5.64 \pm 0.88$ &   $5.40 \pm 0.64$ &   $ -1.32$ &  $ -0.66$ &  $ -0.07$ &  $ 0.48$ & $ 5.14$\\
0149+359(2) &     28.1650 & $    36.1714$ & P & 0.0163 & 81 &  $5.98 \pm 1.36$ &   $5.51 \pm 0.55$ &   \nodata & \nodata & $ -0.15$ &  \nodata & \nodata & $ 3.89$ & \nodata\\
0154+320(1) &     29.3257 & $    32.2400$ & $-1$ & 0.0894 & 372 &  $91.00 \pm 2.00$ &   $40.37 \pm 1.40$ &   $<6.40$ & $<1.67$ & $ -1.46$ &  $< -1.89$ &  \nodata & $ 1.26$ & \nodata\\
0154+320(2) &     29.3175 & $    32.2465$ & C & 0.0894 & 372 &  $91.00 \pm 2.00$ &   $16.98 \pm 0.92$ &   $3.15 \pm 1.08$ &   $<1.59$ & $ -3.02$ &  $ -1.73$ &  $< -1.04$ &  $ 0.53$ & \nodata\\
0304$-$122(1) &     46.7230 & $   -12.1091$ & $-1$ & 0.0788 & 501 &  $86.06 \pm 2.70$ &   $40.07 \pm 1.89$ &   $12.73 \pm 2.30$ &   $2.35 \pm 1.08$ &   $ -1.38$ &  $ -1.18$ &  $ -2.58$ &  $ 1.54$ & $ 0.35$\\
0304$-$122(2) &     46.7184 & $   -12.1057$ & $-1$ & 0.0788 & 501 &  $87.75 \pm 2.00$ &   $45.55 \pm 1.53$ &   $11.67 \pm 1.40$ &   $6.56 \pm 1.35$ &   $ -1.18$ &  $ -1.40$ &  $ -0.88$ &  $ 2.79$ & $ 3.43$\\
0304$-$122(3) &     46.7193 & $   -12.1061$ & C & 0.0788 & 501 &  $87.75 \pm 2.00$ &   $45.55 \pm 1.53$ &   $4.93 \pm 1.44$ &   $2.93 \pm 0.86$ &   $ -1.18$ &  $ -2.28$ &  $ -0.80$ &  $ 2.79$ & $ 1.63$\\
0306$-$237 &     47.0678 & $   -23.5638$ & $-1$ & 0.0665 & 117 &  $55.70 \pm 1.20$ &   $33.71 \pm 0.51$ &   $16.80 \pm 0.59$ &   $14.15 \pm 1.06$ &   $ -0.90$ &  $ -0.71$ &  $ -0.26$ &  $ 3.97$ & $11.67$\\
0431$-$134(1) &     68.5434 & $   -13.3701$ & $-1$ & 0.0327 & 1160 &  $106.00 \pm 3.50$ &   $33.06 \pm 0.98$ &   $23.10 \pm 0.73$ &   $16.96 \pm 0.74$ &   $ -2.10$ &  $ -0.37$ &  $ -0.47$ &  $ 0.23$ & $11.98$\\
0445$-$205 &     72.0125 & $   -20.4440$ & P & 0.0734 & 95 &  $15.50 \pm 2.00$ &   $4.29 \pm 0.52$ &   $2.30 \pm 0.80$ &   $<1.97$ & $ -2.31$ &  $ -0.64$ &  $< -0.24$ &  $ 0.02$ & \nodata\\
0446$-$205 &     72.0431 & $   -20.4160$ & C & 0.0734 & 119 &  $46.90 \pm 1.60$ &   $17.37 \pm 0.52$ &   $10.24 \pm 0.63$ &   $8.58 \pm 0.97$ &   $ -1.79$ &  $ -0.54$ &  $ -0.27$ &  $ 0.25$ & $ 7.03$\\
0717+559 &    110.3399 & $    55.8091$ & $-1$ & 0.0381 & 16 &  $7.16 \pm 1.17$ &   $2.29 \pm 0.64$ &   $<2.12$ & $<1.46$ & $ -2.05$ &  $< -0.08$ &  \nodata & $ 0.02$ & \nodata\\
0810+665 &    123.7207 & $    66.4476$ & $-1$ & 0.1380 & 266 &  $79.90 \pm 1.50$ &   $38.21 \pm 0.66$ &   $9.18 \pm 1.30$ &   $6.17 \pm 0.99$ &   $ -1.33$ &  $ -1.46$ &  $ -0.61$ &  $ 1.65$ & $ 3.95$\\
0816+526(1) &    124.9492 & $    52.5368$ & $-1$ & 0.1890 & 2020 &  $281.00 \pm 2.30$ &   $140.29 \pm 2.33$ &   $48.30 \pm 4.32$ &   $19.62 \pm 1.80$ &   $ -1.25$ &  $ -1.09$ &  $ -1.38$ &  $ 7.27$ & $ 7.12$\\
0816+526(2) &    124.9480 & $    52.5408$ & C & 0.1890 & 2020 &  \nodata & $145.40 \pm 4.38$ &   $23.90 \pm 6.30$ &   $3.59 \pm 1.07$ &   \nodata & $ -1.85$ &  $ -2.90$ &  \nodata & $ 0.43$\\
0816+526(3) &    124.9471 & $    52.5450$ & $-1$ & 0.1890 & 2020 &  $497.10 \pm 2.80$ &   $260.00 \pm 3.00$ &   $59.24 \pm 4.12$ &   $35.33 \pm 2.40$ &   $ -1.17$ &  $ -1.52$ &  $ -0.79$ &  $16.43$ & $19.75$\\
0836+290 &    129.8160 & $    28.8441$ & P & 0.0788 & 1022 &  $156.00 \pm 2.34$ &   $153.60 \pm 0.91$ &   $117.06 \pm 1.25$ &   $95.22 \pm 1.18$ &   $ -0.03$ &  $ -0.28$ &  $ -0.32$ &  $143.78$ & $75.48$\\
0909+162(1) &    138.1463 & $    15.9998$ & C & 0.0851 & 183 &  $25.00 \pm 2.00$ &   $11.57 \pm 1.45$ &   $2.94 \pm 0.83$ &   $1.43 \pm 0.64$ &   $ -1.39$ &  $ -1.41$ &  $ -1.10$ &  $ 0.43$ & $ 0.64$\\
0909+162(2) &    138.1408 & $    15.9958$ & $-1$ & 0.0851 & 183 &  $34.75 \pm 1.35$ &   $18.25 \pm 1.10$ &   $<2.50$ & $<1.89$ & $ -1.16$ &  $< -2.04$ &  \nodata & $ 1.17$ & \nodata\\
0909+161 &    138.1274 & $    15.9244$ & P & 0.0851 & 23 &  $16.41 \pm 0.91$ &   $10.43 \pm 0.34$ &   $6.56 \pm 0.57$ &   $7.56 \pm 0.63$ &   $ -0.82$ &  $ -0.48$ &  $  0.22$ &  $ 1.51$ & $ 8.87$\\
1058+107 &    165.2392 & $    10.5055$ & P & 0.0360 & 24 &  $9.96 \pm 0.39$ &   $7.93 \pm 0.29$ &   $4.85 \pm 0.68$ &   $4.22 \pm 0.74$ &   $ -0.41$ &  $ -0.50$ &  $ -0.21$ &  $ 3.02$ & $ 3.61$\\
1108+289A &    167.6990 & $    28.6601$ & C & 0.0321 & 34 &  $9.50 \pm 1.30$ &   $3.94 \pm 0.72$ &   $2.55 \pm 0.57$ &   $2.59 \pm 1.01$ &   $ -1.59$ &  $ -0.45$ &  $  0.02$ &  $ 0.09$ & $ 2.63$\\
1108+411(1) &    167.9145 & $    40.8380$ & $-1$ & 0.0794 & 771 &  $235.00 \pm 2.00$ &   $67.54 \pm 1.21$ &   $16.30 \pm 2.17$ &   $8.59 \pm 2.57$ &   $ -2.25$ &  $ -1.46$ &  $ -0.98$ &  $ 0.33$ & $ 4.18$\\
1108+411(2) &    167.9144 & $    40.8406$ & C & 0.0794 & 771 &  $235.00 \pm 2.00$ &   $37.70 \pm 0.82$ &   $13.37 \pm 1.65$ &   $3.76 \pm 1.03$ &   $ -3.30$ &  $ -1.06$ &  $ -1.94$ &  $ 0.19$ & $ 0.90$\\
1108+411(3) &    167.9120 & $    40.8391$ & C & 0.0794 & 771 &  $235.00 \pm 2.00$ &   $71.01 \pm 1.46$ &   $10.37 \pm 2.17$ &   \nodata & $ -2.16$ &  $ -1.97$ &  \nodata & $ 0.35$ & \nodata\\
1108+410A & 167.9184 & 40.7853 & $-1$ & 0.0794 & 116 &  $<4.35$ & $<1.21$ & $<2.37$ & $<1.46$ & \nodata & \nodata & \nodata & \nodata & \nodata\\
1108+410B &    167.9319 & $    40.8210$ & P & 0.0794 & 23 &  $6.99 \pm 1.08$ &   $5.27 \pm 0.79$ &   $<2.44$ & $<1.49$ & $ -0.51$ &  $< -0.79$ &  \nodata & $ 1.59$ & \nodata\\
1113+295B(1) &    169.1192 & $    29.2860$ & $-1$ & 0.0471 & 22 &  $15.92 \pm 4.15$ &   $<4.86$ & $4.14 \pm 1.05$ &   $2.16 \pm 0.66$ &   $< -2.14$ &  \nodata & $ -1.00$ &  \nodata & $ 1.04$\\
1113+295B(2) &    169.0945 & $    29.2523$ & P & 0.0471 & 22 &  $54.08 \pm 4.67$ &   $31.78 \pm 4.06$ &   \nodata & \nodata & $ -0.96$ &  \nodata & \nodata & $ 3.30$ & \nodata\\
1113+295C &    169.1438 & $    29.2547$ & C & 0.0471 & 1888 &  $289.90 \pm 6.10$ &   $97.35 \pm 2.90$ &   $35.72 \pm 1.95$ &   $37.28 \pm 1.07$ &   $ -1.97$ &  $ -1.03$ &  $  0.07$ &  $ 0.93$ & $39.12$\\
1129+562 &    173.0960 & $    55.9678$ & $-1$ & 0.0531 & 39 &  $8.22 \pm 0.49$ &   $2.60 \pm 0.59$ &   $<2.00$ & $<2.33$ & $ -2.07$ &  $< -0.27$ &  \nodata & $ 0.02$ & \nodata\\
1130+148 &    173.2561 & $    14.5346$ & C & 0.0834 & 167 &  $13.68 \pm 0.41$ &   $8.82 \pm 0.39$ &   $3.42 \pm 0.76$ &   $3.56 \pm 1.08$ &   $ -0.79$ &  $ -0.97$ &  $  0.06$ &  $ 1.36$ & $ 3.73$\\
1130$-$037 &    173.2713 & $    -4.0133$ & P & 0.0484 & 791 &  $158.51 \pm 2.10$ &   $107.40 \pm 1.37$ &   $62.13 \pm 0.86$ &   \nodata & $ -0.70$ &  $ -0.56$ &  \nodata & $20.45$ & \nodata\\
1131+493 &    173.4967 & $    49.0622$ & C & 0.0338 & 835 &  $131.25 \pm 2.00$ &   $91.44 \pm 1.32$ &   $52.85 \pm 1.10$ &   $41.84 \pm 0.92$ &   $ -0.65$ &  $ -0.56$ &  $ -0.36$ &  $19.60$ & $32.17$\\
1132+492 &    173.6940 & $    48.9562$ & C & 0.0338 & 475 &  $30.38 \pm 0.83$ &   $31.86 \pm 0.74$ &   $26.76 \pm 0.88$ &   $18.59 \pm 0.86$ &   $  0.09$ &  $ -0.18$ &  $ -0.56$ &  $39.02$ & $12.34$\\
1132+493 &    173.7056 & $    49.0779$ & P & 0.0338 & 31 &  $12.61 \pm 0.49$ &   $7.15 \pm 0.43$ &   $<2.45$ & $1.81 \pm 0.83$ &   $ -1.02$ &  $< -1.10$ &  \nodata & $ 0.64$ & \nodata\\
1141+466(1) &    175.9146 & $    46.3549$ & P & 0.1162 & 814 &  $149.40 \pm 0.50$ &   $44.90 \pm 0.66$ &   $3.47 \pm 1.04$ &   $1.31 \pm 0.73$ &   $ -2.17$ &  $ -2.63$ &  $ -1.49$ &  $ 0.27$ & $ 0.44$\\
1141+466(2) &    175.9158 & $    46.3564$ & C & 0.1162 & 814 &  $149.40 \pm 0.50$ &   $66.15 \pm 0.62$ &   $11.35 \pm 1.40$ &   $0.95 \pm 0.56$ &   $ -1.47$ &  $ -1.81$ &  $ -3.79$ &  $ 0.39$ & $ 0.06$\\
1141+676 &    176.1526 & $    67.4060$ & P & 0.1164 & 196 &  $68.94 \pm 0.47$ &   $37.75 \pm 0.51$ &   $16.84 \pm 1.20$ &   $10.48 \pm 0.95$ &   $ -1.08$ &  $ -0.83$ &  $ -0.72$ &  $ 2.90$ & $ 6.15$\\
1142+198 &    176.2709 & $    19.6064$ & C & 0.0214 & 5450 &  $723.75 \pm 9.55$ &   $432.27 \pm 4.22$ &   $215.48 \pm 2.70$ &   $172.11 \pm 2.30$ &   $ -0.93$ &  $ -0.71$ &  $ -0.34$ &  $48.08$ & $133.66$\\
1153+736 &    178.9965 & $    73.4154$ & P & 0.0836 & 64 &  $27.66 \pm 0.59$ &   $19.33 \pm 0.75$ &   $3.01 \pm 1.09$ &   $3.96 \pm 1.74$ &   $ -0.65$ &  $ -1.91$ &  $  0.42$ &  $ 4.20$ & $ 5.39$\\
1155+266 &    179.5839 & $    26.3533$ & C & 0.1120 & 880 &  $13.80 \pm 6.70$ &   $1.63 \pm 0.73$ &   $6.92 \pm 0.93$ &   $7.09 \pm 0.64$ &   $ -3.85$ &  $  1.48$ &  $  0.04$ &  $ 0.00$ & $ 7.29$\\
1159+583(1) &    180.5143 & $    58.0337$ & $-1$ & 0.1035 & 765 &  $110.00 \pm 1.00$ &   $53.15 \pm 0.88$ &   $19.66 \pm 2.33$ &   $8.70 \pm 1.70$ &   $ -1.31$ &  $ -1.02$ &  $ -1.25$ &  $ 2.40$ & $ 3.48$\\
1159+583(3) &    180.5203 & $    58.0373$ & $-1$ & 0.1035 & 765 &  $124.00 \pm 1.10$ &   $54.62 \pm 0.86$ &   $23.74 \pm 3.00$ &   $12.50 \pm 2.10$ &   $ -1.48$ &  $ -0.85$ &  $ -0.98$ &  $ 1.66$ & $ 6.07$\\
1201+282 &    180.9028 & $    27.9443$ & C & 0.1390 & 215 &  $2.93 \pm 0.31$ &   $1.87 \pm 0.43$ &   $4.94 \pm 0.86$ &   $1.69 \pm 0.74$ &   $ -0.81$ &  $  1.00$ &  $ -1.64$ &  $ 0.28$ & $ 0.51$\\
1201+026(1) &    181.0303 & $     2.4099$ & C & 0.0844 & 244 &  $120.00 \pm 1.00$ &   $29.82 \pm 1.32$ &   $3.75 \pm 1.50$ &   \nodata & $ -2.51$ &  $ -2.13$ &  \nodata & $ 0.26$ & \nodata\\
1201+026(2) &    181.0264 & $     2.4118$ & C & 0.0844 & 244 &  $120.00 \pm 1.00$ &   $39.40 \pm 1.40$ &   $8.11 \pm 0.96$ &   $6.06 \pm 0.90$ &   $ -2.01$ &  $ -1.62$ &  $ -0.45$ &  $ 0.34$ & $ 4.37$\\
1201+026(3) &    181.2723 & $     2.4135$ & C & 0.0844 & 244 &  $120.00 \pm 1.00$ &   $25.65 \pm 1.06$ &   \nodata & \nodata & $ -2.78$ &  \nodata & \nodata & $ 0.22$ & \nodata\\
1207+722 &    182.5808 & $    71.9993$ & C & 0.1226 & 256 &  \nodata & $47.53 \pm 2.40$ &   $6.56 \pm 2.09$ &   $<2.58$ & \nodata & $ -2.03$ &  $< -1.43$ &  \nodata & \nodata\\
1221+615(1) &    185.8762 & $    61.2473$ & $-1$ & 0.2308 & 321 &  $69.15 \pm 0.57$ &   $34.34 \pm 0.89$ &   $12.50 \pm 1.49$ &   $3.89 \pm 1.35$ &   $ -1.26$ &  $ -1.04$ &  $ -1.78$ &  $ 1.74$ & $ 1.05$\\
1221+615(2) &    185.8740 & $    61.2521$ & $-1$ & 0.2308 & 321 &  $49.46 \pm 0.54$ &   $24.78 \pm 0.69$ &   $5.98 \pm 1.55$ &   $<2.58$ & $ -1.24$ &  $ -1.46$ &  $< -1.28$ &  $ 1.30$ & \nodata\\
1224+091 &    186.8264 & $     8.8431$ & C & 0.0896 & 48 &  $<3.40$ & $5.05 \pm 0.68$ &   $<2.39$ & $<2.91$ & \nodata & $< -0.77$ &  \nodata & \nodata & \nodata\\
1225+636(1) &    186.9687 & $    63.3840$ & $-1$ & 0.1459 & 210 &  $70.32 \pm 0.82$ &   $19.53 \pm 0.65$ &   $11.80 \pm 4.80$ &   $<1.66$ & $ -2.31$ &  $ -0.52$ &  $< -3.00$ &  $ 0.08$ & \nodata\\
1225+636(2) &    186.9634 & $    63.3848$ & C & 0.1459 & 210 &  $70.32 \pm 0.82$ &   $22.41 \pm 0.63$ &   $5.04 \pm 1.27$ &   $5.47 \pm 1.00$ &   $ -2.06$ &  $ -1.53$ &  $  0.13$ &  $ 0.10$ & $ 6.00$\\
1231+674 &    188.3085 & $    67.1289$ & C & 0.1071 & 879 &  $18.00 \pm 1.00$ &   $11.00 \pm 1.33$ &   $7.33 \pm 1.23$ &   $4.43 \pm 0.84$ &   $ -0.89$ &  $ -0.42$ &  $ -0.77$ &  $ 1.35$ & $ 2.51$\\
1232+414(1) &    188.6250 & $    41.1599$ & C & 0.1908 & 689 &  $<17.00$ & $19.51 \pm 2.40$ &   $<2.67$ & $<1.39$ & \nodata & $< -2.04$ &  \nodata & \nodata & \nodata\\
1232+414(2) &    188.6142 & $    41.1668$ & $-1$ & 0.1908 & 689 &  $105.30 \pm 1.50$ &   $48.94 \pm 0.71$ &   $19.85 \pm 2.50$ &   $12.67 \pm 1.35$ &   $ -1.38$ &  $ -0.93$ &  $ -0.69$ &  $ 1.87$ & $ 7.65$\\
1233+169 &    189.0338 & $    16.6414$ & P & 0.0784 & 630 &  $189.00 \pm 2.50$ &   $64.22 \pm 1.16$ &   $33.21 \pm 0.99$ &   $20.39 \pm 0.96$ &   $ -1.94$ &  $ -0.68$ &  $ -0.75$ &  $ 0.65$ & $11.78$\\
1233+168 &    189.1079 & $    16.5384$ & C & 0.0784 & 1338 &  $65.00 \pm 12.00$ &   $11.13 \pm 2.40$ &   $4.62 \pm 1.17$ &   $3.28 \pm 0.90$ &   $ -3.18$ &  $ -0.90$ &  $ -0.52$ &  $ 0.01$ & $ 2.23$\\
1238+188 &    190.2511 & $    18.5537$ & C & 0.0718 & 537 &  $48.35 \pm 1.41$ &   $18.32 \pm 0.80$ &   $5.77 \pm 1.03$ &   $4.75 \pm 0.91$ &   $ -1.75$ &  $ -1.19$ &  $ -0.30$ &  $ 0.29$ & $ 3.82$\\
1243+699 &    191.4726 & $    69.6582$ & C & 0.2307 & 220 &  $38.76 \pm 1.39$ &   $6.66 \pm 0.60$ &   $4.27 \pm 1.26$ &   $0.80 \pm 0.43$ &   $ -3.17$ &  $ -0.46$ &  $ -2.56$ &  $ 0.00$ & $ 0.12$\\
1256+281 &    194.8462 & $    27.9112$ & C & 0.0231 & 450 &  $59.69 \pm 1.39$ &   $37.62 \pm 1.53$ &   $<2.56$ & $<1.85$ & $ -0.83$ &  $< -2.76$ &  \nodata & $ 5.26$ & \nodata\\
1257+282(1) &    194.8964 & $    27.9579$ & $-1$ & 0.0231 & 215 &  $39.83 \pm 0.89$ &   $22.64 \pm 0.45$ &   $7.83 \pm 1.94$ &   $3.00 \pm -1.00$ &   $ -1.02$ &  $ -1.09$ &  $ -1.47$ &  $ 2.04$ & $ 1.02$\\
1257+282(2) &    194.9002 & $    27.9610$ & $-1$ & 0.0231 & 215 &  $50.20 \pm 1.00$ &   $23.19 \pm 0.47$ &   $7.54 \pm 1.18$ &   $2.82 \pm 0.71$ &   $ -1.39$ &  $ -1.15$ &  $ -1.50$ &  $ 0.86$ & $ 0.93$\\
1301+195 &    195.9442 & $    19.2715$ & P & 0.0649 & 74 &  $22.51 \pm 0.28$ &   $13.70 \pm 0.27$ &   $6.20 \pm 0.75$ &   $1.96 \pm 0.55$ &   $ -0.89$ &  $ -0.81$ &  $ -1.76$ &  $ 1.65$ & $ 0.54$\\
1300+677 &    195.6686 & $    67.4780$ & $-1$ & 0.1055 & 298 &  $137.20 \pm 0.60$ &   $87.48 \pm 0.77$ &   $15.90 \pm 0.90$ &   $<3.00$ & $ -0.81$ &  $ -1.75$ &  $< -2.55$ &  $12.86$ & \nodata\\
1320+584(1) &    200.7321 & $    58.1675$ & $-1$ & 0.1932 & 325 &  $78.62 \pm 1.02$ &   $50.00 \pm 1.00$ &   $13.94 \pm 1.33$ &   \nodata & $ -0.82$ &  $ -1.31$ &  \nodata & $ 7.27$ & \nodata\\
1320+584(2) &    200.7268 & $    58.1711$ & C & 0.1932 & 325 &  $53.21 \pm 0.81$ &   $27.87 \pm 0.67$ &   $<2.85$ & \nodata & $ -1.16$ &  $< -2.34$ &  \nodata & $ 1.77$ & \nodata\\
1333+412(1) &    203.8350 & $    40.9999$ & $-1$ & 0.2290 & 797 &  $169.70 \pm 0.90$ &   $63.05 \pm 0.62$ &   $22.77 \pm 1.40$ &   \nodata & $ -1.78$ &  $ -1.04$ &  \nodata & $ 0.93$ & \nodata\\
1333+412(2) &    203.8320 & $    41.0024$ & $-1$ & 0.2290 & 797 &  $161.30 \pm 0.90$ &   $62.34 \pm 0.64$ &   $18.44 \pm 1.21$ &   \nodata & $ -1.71$ &  $ -1.25$ &  \nodata & $ 1.09$ & \nodata\\
1339+266A &    205.4552 & $    26.3738$ & $-1$ & 0.0724 & 40 &  $15.90 \pm 0.64$ &   $6.39 \pm 0.67$ &   $<1.63$ & $<1.59$ & $ -1.64$ &  $< -1.40$ &  \nodata & $ 0.13$ & \nodata\\
1339+266B &    205.4606 & $    26.3715$ & $-1$ & 0.0724 & 287 &  $91.95 \pm 0.61$ &   $41.04 \pm 0.64$ &   $1.87 \pm -1.00$ &   $<1.59$ & $ -1.45$ &  $ -3.17$ &  $< -0.25$ &  $ 1.32$ & \nodata\\
1346+268A &    207.2186 & $    26.5928$ & P & 0.0622 & 883 &  $234.60 \pm 0.50$ &   $110.74 \pm 0.84$ &   $21.79 \pm 1.02$ &   $12.24 \pm 0.67$ &   $ -1.35$ &  $ -1.67$ &  $ -0.88$ &  $ 4.52$ & $ 6.40$\\
1346+268B &    207.2474 & $    26.5594$ & $-1$ & 0.0622 & 35 &  $10.75 \pm 0.67$ &   $12.05 \pm 1.17$ &   $4.14 \pm 0.96$ &   $<1.50$ & $  0.21$ &  $ -1.10$ &  $< -1.55$ &  $19.60$ & \nodata\\
1415+084(1) &    214.3803 & $     8.2084$ & C & 0.0570 & 331 &  $29.83 \pm 1.66$ &   $8.27 \pm 0.72$ &   $3.60 \pm 0.59$ &   \nodata & $ -2.31$ &  $ -0.85$ &  \nodata & $ 0.03$ & \nodata\\
1415+084(2) &    214.3826 & $     8.2101$ & C & 0.0570 & 331 &  $29.83 \pm 1.66$ &   $5.88 \pm 0.82$ &   $<0.78$ & \nodata & $ -2.93$ &  $< -2.07$ &  \nodata & $ 0.02$ & \nodata\\
1418+253(1) &    215.1731 & $    25.1499$ & $-1$ & 0.0780 & 116 &  $22.27 \pm 0.51$ &   $9.45 \pm 0.75$ &   $1.28 \pm 0.50$ &   $1.70 \pm 0.71$ &   $ -1.54$ &  $ -2.05$ &  $  0.43$ &  $ 0.24$ & $ 2.34$\\
1418+253(2) &    215.1747 & $    25.1461$ & $-1$ & 0.0780 & 116 &  $<26.70$ & $6.63 \pm 0.76$ &   $2.03 \pm 0.66$ &   $3.76 \pm 0.84$ &   \nodata & $ -1.21$ &  $  0.94$ &  \nodata & $ 7.53$\\
1418+253(3) &    215.1758 & $    25.1438$ & C & 0.0780 & 116 &  $<26.70$ & $6.65 \pm 0.79$ &   $2.15 \pm 0.80$ &   $<1.56$ & $ -2.93$ &  $ -1.16$ &  $< -0.49$ &  \nodata & \nodata\\
1418+253(4) &    215.1767 & $    25.1408$ & $-1$ & 0.0780 & 116 &  $21.03 \pm 0.59$ &   $6.36 \pm 0.65$ &   $<1.74$ & $<1.84$ & $ -2.15$ &  $< -1.33$ &  \nodata & $ 0.04$ & \nodata\\
1424+169(1) &    216.6422 & $    16.7507$ & $-1$ & 0.0528 & 97 &  $9.55 \pm 0.58$ &   $8.20 \pm 0.73$ &   \nodata & \nodata & $ -0.27$ &  \nodata & \nodata & $ 4.28$ & \nodata\\
1424+169(2) &    216.6313 & $    16.7633$ & $-1$ & 0.0528 & 97 &  $7.94 \pm 0.70$ &   $1.92 \pm 0.63$ &   \nodata & \nodata & $ -2.56$ &  \nodata & \nodata & $ 0.00$ & \nodata\\
1424+167(1) &    216.8221 & $    16.5548$ & $-1$ & 0.0528 & 103 &  $11.93 \pm 0.57$ &   $7.15 \pm 0.44$ &   $4.75 \pm 1.32$ &   \nodata & $ -0.92$ &  $ -0.42$ &  \nodata & $ 0.81$ & \nodata\\
1435+249(1) &    219.3126 & $    24.7591$ & C & 0.0883 & 175 &  $11.80 \pm 0.56$ &   $9.21 \pm 0.39$ &   $7.72 \pm 0.61$ &   $5.77 \pm 0.58$ &   $ -0.45$ &  $ -0.18$ &  $ -0.45$ &  $ 3.20$ & $ 4.16$\\
1435+249(2) &    219.3168 & $    24.7653$ & $-1$ & 0.0883 & 175 &  $11.69 \pm 0.88$ &   $5.54 \pm 0.71$ &   $<1.98$ & \nodata & $ -1.35$ &  $< -1.06$ &  \nodata & $ 0.23$ & \nodata\\
1435+250 &    219.3200 & $    24.8693$ & C & 0.0883 & 206 &  $23.03 \pm 0.73$ &   $12.22 \pm 0.45$ &   $6.53 \pm 1.10$ &   $4.62 \pm 0.83$ &   $ -1.14$ &  $ -0.64$ &  $ -0.53$ &  $ 0.82$ & $ 3.12$\\
1433+553 &    218.8688 & $    55.1311$ & C & 0.1396 & 447 &  $73.92 \pm 1.02$ &   $16.81 \pm 0.55$ &   $7.95 \pm 0.83$ &   \nodata & $ -2.67$ &  $ -0.77$ &  \nodata & $ 0.03$ & \nodata\\
1435+038(0) &    219.5993 & $     3.6729$ & C & 0.2240 & 801 &  $103.16 \pm 0.78$ &   $11.25 \pm 0.85$ &   $<16.06$ & \nodata & $ -3.99$ &  $<  0.37$ &  \nodata & $ 0.41$ & \nodata\\
1435+038(1) &    219.5946 & $     3.6713$ & $-1$ & 0.2240 & 801 &  $103.16 \pm 0.78$ &   $47.29 \pm 0.54$ &   $12.21 \pm 2.13$ &   \nodata & $ -1.40$ &  $ -1.39$ &  \nodata & $ 1.70$ & \nodata\\
1435+038(2) &    219.5886 & $     3.6702$ & $-1$ & 0.2240 & 801 &  $114.12 \pm 0.73$ &   $55.45 \pm 0.51$ &   $15.91 \pm 1.61$ &   \nodata & $ -1.30$ &  $ -1.28$ &  \nodata & $ 2.56$ & \nodata\\
1435+038(4) &    219.5960 & $     3.6714$ & $-1$ & 0.2240 & 801 &  $89.03 \pm 1.37$ &   $<2.54$ & $<2.93$ & \nodata & $< -6.41$ &  \nodata & \nodata & \nodata & \nodata\\
1452+188 &    223.6312 & $    18.6423$ & P & 0.0579 & 38 &  $11.99 \pm 0.23$ &   $7.71 \pm 0.25$ &   $2.11 \pm 0.56$ &   \nodata & $ -0.80$ &  $ -1.33$ &  \nodata & $ 1.17$ & \nodata\\
1508+059(1) &    227.7339 & $     5.7446$ & $-1$ & 0.0767 & 489 &  $60.51 \pm 0.46$ &   $21.31 \pm 0.42$ &   $4.07 \pm 1.03$ &   \nodata & $ -1.88$ &  $ -1.70$ &  \nodata & $ 0.25$ & \nodata\\
1508+065(1) &    227.8630 & $     6.3472$ & $-1$ & 0.0817 & 552 &  $127.65 \pm 0.72$ &   $54.79 \pm 0.66$ &   $18.02 \pm -1.00$ &   \nodata & $ -1.52$ &  $ -1.14$ &  \nodata & $ 1.49$ & \nodata\\
1508+065(2) &    227.8584 & $     6.3503$ & $-1$ & 0.0817 & 552 &  $127.44 \pm 0.70$ &   $56.48 \pm 0.62$ &   $15.35 \pm 2.17$ &   \nodata & $ -1.47$ &  $ -1.34$ &  \nodata & $ 1.76$ & \nodata\\
1508+182(1) &    227.7893 & $    18.0300$ & $-1$ & 0.1163 & 346 &  $74.34 \pm 0.88$ &   $30.50 \pm 0.65$ &   $<9.33$ & \nodata & $ -1.60$ &  $< -1.22$ &  \nodata & $ 0.68$ & \nodata\\
1508+182(2) &    227.7847 & $    18.0326$ & $-1$ & 0.1163 & 346 &  $68.61 \pm 0.96$ &   $26.85 \pm 0.61$ &   $6.90 \pm 1.51$ &   \nodata & $ -1.69$ &  $ -1.39$ &  \nodata & $ 0.49$ & \nodata\\
1510+076(1) &    228.1405 & $     7.4249$ & $-1$ & 0.0451 & 17 &  $10.59 \pm 1.54$ &   $2.76 \pm 0.55$ &   $<2.87$ & \nodata & $ -2.42$ &  $<  0.04$ &  \nodata & $ 0.01$ & \nodata\\
1510+076(2) &    228.1427 & $     7.4318$ & C & 0.0451 & 17 &  $10.59 \pm 1.54$ &   $1.97 \pm 0.45$ &   $1.57 \pm 0.70$ &   \nodata & $ -3.03$ &  $ -0.23$ &  \nodata & $ 0.01$ & \nodata\\
1514+072 &    229.1854 & $     7.0216$ & P & 0.0348 & 5390 &  $897.61 \pm 2.89$ &   $691.40 \pm 2.70$ &   $393.16 \pm 5.70$ &   \nodata & $ -0.47$ &  $ -0.58$ &  \nodata & $227.36$ & \nodata\\
1520+087 &    230.7719 & $     8.6094$ & $-1$ & 0.0355 & 13 &  $5.59 \pm 0.47$ &   $2.11 \pm 0.37$ &   $<2.30$ & \nodata & $ -1.75$ &  $<  0.09$ &  \nodata & $ 0.03$ & \nodata\\
1525+290 &    231.9350 & $    28.9183$ & C & 0.0656 & 224 &  $88.69 \pm 0.64$ &   $28.84 \pm 0.78$ &   $5.70 \pm 0.84$ &   $3.31 \pm 0.90$ &   $ -2.02$ &  $ -1.66$ &  $ -0.83$ &  $ 0.24$ & $ 1.80$\\
1530+282 &    233.1860 & $    28.0631$ & C & 0.0734 & 352 &  $133.24 \pm 1.03$ &   $14.71 \pm 0.69$ &   $4.87 \pm 0.82$ &   $1.99 \pm 1.13$ &   $ -3.97$ &  $ -1.13$ &  $ -1.37$ &  $ 0.00$ & $ 0.73$\\
1531+312(1) &    233.3132 & $    31.1285$ & $-1$ & 0.0670 & 49 &  $6.89 \pm 0.37$ &   $2.81 \pm 0.32$ &   $<2.17$ & $<1.61$ & $ -1.62$ &  $< -0.27$ &  \nodata & $ 0.06$ & \nodata\\
1531+312(2) &    233.3137 & $    31.1307$ & C & 0.0670 & 49 &  $6.89 \pm 0.37$ &   $1.28 \pm 0.32$ &   $<2.17$ & $<2.80$ & $ -3.03$ &  $<  0.54$ &  \nodata & $ 0.03$ & \nodata\\
1531+312(3) &    233.3169 & $    31.1331$ & $-1$ & 0.0670 & 49 &  $8.53 \pm 0.38$ &   $4.39 \pm 0.45$ &   $<2.17$ & $<2.25$ & $ -1.20$ &  $< -0.72$ &  \nodata & $ 0.26$ & \nodata\\
1555+356(2) &    239.4258 & $    35.5076$ & $-1$ & 0.1579 & 216 &  $57.62 \pm 0.63$ &   $14.60 \pm 0.41$ &   $7.02 \pm 1.85$ &   $3.03 \pm 0.90$ &   $ -2.47$ &  $ -0.75$ &  $ -1.28$ &  $ 0.04$ & $ 1.18$\\
1555+356(3) &    239.4267 & $    35.5094$ & C & 0.1579 & 216 &  $23.70 \pm 0.44$ &   $16.80 \pm 0.44$ &   $5.38 \pm 0.91$ &   $3.58 \pm 1.04$ &   $ -0.62$ &  $ -1.17$ &  $ -0.62$ &  $ 0.83$ & $ 2.26$\\
1556+274 &    239.5585 & $    27.2723$ & $-1$ & 0.0896 & 130 &  $30.38 \pm 0.41$ &   $16.97 \pm 0.36$ &   $7.97 \pm 1.59$ &   $<1.92$ & $ -1.05$ &  $ -0.77$ &  $< -2.18$ &  $ 1.42$ & \nodata\\
1559+161(1) &    240.5703 & $    15.9745$ & $-1$ & 0.0354 & 17 &  $6.83 \pm 0.36$ &   $4.97 \pm 0.42$ &   $<1.80$ & \nodata & $ -0.57$ &  $< -1.04$ &  \nodata & $ 1.28$ & \nodata\\
1602+178B(1) &    241.2872 & $    17.7314$ & $-1$ & 0.0368 & 780 &  $116.48 \pm 1.62$ &   $76.02 \pm 2.36$ &   $10.84 \pm 1.27$ &   \nodata & $ -0.77$ &  $ -2.00$ &  \nodata & $12.34$ & \nodata\\
1602+178B(1.5) &    241.2875 & $    17.7299$ & $-1$ & 0.0368 & 780 &  \nodata & \nodata & $10.75 \pm 1.26$ &   \nodata & \nodata & \nodata & \nodata & \nodata & \nodata\\
1602+178B(2) &    241.2880 & $    17.7270$ & $-1$ & 0.0368 & 780 &  $213.70 \pm 1.75$ &   $139.80 \pm 2.39$ &   $<4.73$ & \nodata & $ -0.76$ &  $< -3.47$ &  \nodata & $22.92$ & \nodata\\
1603+165 &    241.3717 & $    16.4357$ & C & 0.0372 & 44 &  $9.04 \pm 0.35$ &   $4.78 \pm 0.33$ &   $2.48 \pm 0.69$ &   \nodata & $ -1.15$ &  $ -0.67$ &  \nodata & $ 0.32$ & \nodata\\
1610+296(2) &    243.1456 & $    29.4814$ & C & 0.0320 & 119 &  $38.00 \pm -1.00$ &   $<0.59$ & $2.83 \pm 0.86$ &   $<1.66$ & $< -7.50$ &  \nodata & $< -0.82$ &  \nodata & \nodata\\
1626+396 &    247.1594 & $    39.5513$ & $-1$ & 0.0299 & 3480 &  $440.00 \pm -1.00$ &   $185.00 \pm -1.00$ &   \nodata & \nodata & $ -1.56$ &  \nodata & \nodata & $ 4.61$ & \nodata\\
1638+468 &    250.0925 & $    46.7131$ & $-1$ & 0.2070 & 212 &  $74.10 \pm 0.73$ &   $44.70 \pm 0.84$ &   \nodata & \nodata & $ -0.91$ &  \nodata & \nodata & $ 5.19$ & \nodata\\
1657+325A &    254.7545 & $    32.4941$ & $-1$ & 0.0628 & 171 &  $22.00 \pm -1.00$ &   $1.18 \pm 0.30$ &   $<3.28$ & $<1.82$ & $ -5.27$ &  $<  1.05$ &  \nodata & $ 0.00$ & \nodata\\
1657+325B & 254.7852 & 32.5000 & $-1$ & 0.0628 & 12 &  \nodata & $<0.62$ & $<2.86$ & \nodata & \nodata & \nodata & \nodata & \nodata & \nodata\\
1707+344(1) &    257.4131 & $    34.4283$ & $-1$ & 0.0806 & 680 &  $100.38 \pm 0.81$ &   $56.81 \pm 1.53$ &   $22.00 \pm 3.20$ &   $13.90 \pm -1.00$ &   $ -1.03$ &  $ -0.97$ &  $ -0.70$ &  $ 5.02$ & $ 8.29$\\
1707+344(2) &    257.4096 & $    34.4336$ & $-1$ & 0.0806 & 680 &  $108.22 \pm 0.74$ &   $62.07 \pm 1.29$ &   $23.00 \pm 2.50$ &   $7.00 \pm 2.80$ &   $ -1.00$ &  $ -1.02$ &  $ -1.82$ &  $ 5.81$ & $ 1.84$\\
1708+345 &    257.4972 & $    34.5120$ & P & 0.0806 & 16 &  $9.53 \pm 0.67$ &   $4.95 \pm 0.81$ &   $3.83 \pm 1.77$ &   $<2.19$ & $ -1.18$ &  $ -0.26$ &  $< -0.85$ &  $ 0.30$ & \nodata\\
1709+397B &    257.7364 & $    39.6926$ & C & 0.0656 & 543 &  $106.15 \pm 0.70$ &   $60.28 \pm 0.75$ &   $23.20 \pm 2.50$ &   $12.13 \pm 2.50$ &   $ -1.02$ &  $ -0.98$ &  $ -0.99$ &  $ 5.41$ & $ 5.85$\\
1712+640 &    258.0974 & $    64.0334$ & C & 0.0808 & 290 &  $82.50 \pm 1.00$ &   $45.76 \pm 1.15$ &   $4.67 \pm 0.77$ &   $<1.77$ & $ -1.06$ &  $ -2.34$ &  $< -1.48$ &  $ 3.71$ & \nodata\\
1712+641(1) &    258.2699 & $    64.1157$ & $-1$ & 0.0808 & 66 &  $26.93 \pm 0.99$ &   $8.41 \pm 0.80$ &   $3.59 \pm 1.23$ &   $<1.87$ & $ -2.10$ &  $ -0.87$ &  $< -1.00$ &  $ 0.06$ & \nodata\\
1712+641(2) &    258.2661 & $    64.1177$ & C & 0.0808 & 66 &  $26.93 \pm 0.99$ &   $8.80 \pm 0.54$ &   $1.28 \pm 0.41$ &   $<1.87$ & $ -2.01$ &  $ -1.98$ &  $<  0.58$ &  $ 0.06$ & \nodata\\
1713+641(1) &    258.3701 & $    64.0443$ & $-1$ & 0.0808 & 250 &  $85.10 \pm 0.66$ &   $37.21 \pm 0.56$ &   $9.33 \pm 1.44$ &   $4.84 \pm 1.30$ &   $ -1.49$ &  $ -1.42$ &  $ -1.00$ &  $ 1.10$ & $ 2.31$\\
1713+641(2) &    258.3706 & $    64.0457$ & C & 0.0808 & 250 &  $85.10 \pm 0.66$ &   $48.83 \pm 0.81$ &   $7.26 \pm 0.79$ &   $5.02 \pm 0.79$ &   $ -1.00$ &  $ -1.96$ &  $ -0.56$ &  $ 1.44$ & $ 3.31$\\
1713+641(3) &    258.3735 & $    64.0503$ & $-1$ & 0.0808 & 250 &  $70.30 \pm 0.65$ &   $37.34 \pm 0.67$ &   $4.26 \pm 0.88$ &   $<1.86$ & $ -1.14$ &  $ -2.23$ &  $< -1.27$ &  $ 2.52$ & \nodata\\
1706+786 &    255.8676 & $    78.6321$ & $-1$ & 0.0581 & 157 &  $42.81 \pm 0.69$ &   $23.50 \pm 0.58$ &   $5.98 \pm 1.11$ &   $<1.78$ & $ -1.08$ &  $ -1.40$ &  $< -1.85$ &  $ 1.82$ & \nodata\\
1705+786 &    255.7604 & $    78.5992$ & $-1$ & 0.0581 & 62 &  $26.00 \pm -1.00$ &   $6.91 \pm 0.73$ &   $<2.30$ & $<1.83$ & $ -2.39$ &  $< -1.13$ &  \nodata & $ 0.02$ & \nodata\\
1706+787 &    255.8668 & $    78.6660$ & P & 0.0581 & 39 &  $5.06 \pm 0.60$ &   $3.96 \pm 0.53$ &   $<1.01$ & $<1.99$ & $ -0.44$ &  $< -1.40$ &  \nodata & $ 1.39$ & \nodata\\
1703+787 &    255.2176 & $    78.6901$ & $-1$ & 0.0581 & 10 &  $2.14 \pm 0.45$ &   $1.25 \pm 0.36$ &   $<2.32$ & $<3.48$ & $ -0.97$ &  $<  0.63$ &  \nodata & $ 0.13$ & \nodata\\
1820+689 &    274.9260 & $    68.9476$ & C/P & 0.0880 & 801 &  $83.27 \pm 1.61$ &   $52.65 \pm 1.54$ &   $31.11 \pm 1.12$ &   $24.36 \pm 0.78$ &   $ -0.83$ &  $ -0.54$ &  $ -0.37$ &  $ 7.47$ & $18.50$\\
1826+747(1) &    276.2349 & $    74.7308$ & $-1$ & 0.1271 & 244 &  $38.53 \pm 1.11$ &   $21.63 \pm 0.71$ &   $1.44 \pm 0.65$ &   $1.85 \pm 0.65$ &   $ -1.04$ &  $ -2.78$ &  $  0.38$ &  $ 1.85$ & $ 2.46$\\
1826+747(2) &    276.2153 & $    74.7316$ & $-1$ & 0.1271 & 244 &  $44.98 \pm 1.27$ &   $16.27 \pm 0.54$ &   $3.68 \pm 1.30$ &   $2.14 \pm 0.76$ &   $ -1.83$ &  $ -1.53$ &  $ -0.83$ &  $ 0.21$ & $ 1.16$\\
1849+702(1) &    282.3256 & $    70.3535$ & C & 0.0899 & 163 &  $18.83 \pm 0.54$ &   $14.47 \pm 0.34$ &   $8.28 \pm 0.49$ &   $9.48 \pm 0.65$ &   $ -0.47$ &  $ -0.57$ &  $  0.21$ &  $ 4.71$ & $11.04$\\
1857+799 &    283.4678 & $    80.0474$ & P & 0.2139 & 180 &  $2.80 \pm 0.51$ &   $2.66 \pm 0.27$ &   $5.34 \pm 1.44$ &   $2.61 \pm 0.60$ &   $ -0.09$ &  $  0.71$ &  $ -1.09$ &  $ 2.15$ & $ 1.17$\\
2124$-$124(1) &    321.7414 & $   -12.2154$ & $-1$ & 0.1760 & 251 &  $96.09 \pm 1.10$ &   $38.45 \pm 1.10$ &   $<2.43$ & $<2.10$ & $ -1.65$ &  $< -2.83$ &  \nodata & $ 0.78$ & \nodata\\
2124$-$124(2) &    321.7384 & $   -12.2144$ & C & 0.1760 & 251 &  $96.09 \pm 1.10$ &   $26.53 \pm 0.58$ &   $8.96 \pm 1.13$ &   $8.38 \pm 0.81$ &   $ -2.32$ &  $ -1.11$ &  $ -0.10$ &  $ 0.54$ & $ 7.77$\\
2142$-$202 &    326.3143 & $   -19.9952$ & $-1$ & 0.0576 & 351 &  $2.45 \pm 1.27$ &   $<0.70$ & $2.28 \pm 1.06$ &   $<1.74$ & $< -2.26$ &  \nodata & $< -0.41$ &  \nodata & \nodata\\
2149$-$158C(1) &    327.9990 & $   -15.6384$ & C & 0.0646 & 176 &  $29.88 \pm 1.80$ &   $12.76 \pm 1.03$ &   $5.24 \pm 1.07$ &   $1.83 \pm 0.75$ &   $ -1.53$ &  $ -0.91$ &  $ -1.61$ &  $ 0.34$ & $ 0.56$\\
2149$-$158C(2) &    327.9796 & $   -15.6263$ & $-1$ & 0.0646 & 176 &  $64.00 \pm 3.40$ &   $14.67 \pm 1.70$ &   $<2.93$ & \nodata & $ -2.65$ &  $< -1.65$ &  \nodata & $ 0.03$ & \nodata\\
2154$-$080A(1) &    329.2532 & $    -7.8474$ & $-1$ & 0.0584 & 460 &  $81.60 \pm 1.14$ &   $51.47 \pm 2.65$ &   $36.00 \pm -1.00$ &   \nodata & $ -0.83$ &  $ -0.37$ &  \nodata & $ 7.22$ & \nodata\\
2154$-$080A(2) &    329.2569 & $    -7.8398$ & C & 0.0584 & 460 &  $81.20 \pm 1.13$ &   $11.11 \pm 1.05$ &   $5.45 \pm 0.53$ &   $4.04 \pm 0.71$ &   $ -3.58$ &  $ -0.73$ &  $ -0.46$ &  $ 0.00$ & $ 2.88$\\
2154$-$080A(3) &    329.2635 & $    -7.8362$ & $-1$ & 0.0584 & 460 &  $73.43 \pm 1.06$ &   $46.53 \pm 2.41$ &   $18.00 \pm -1.00$ &   \nodata & $ -0.82$ &  $ -0.97$ &  \nodata & $ 6.66$ & \nodata\\
2154$-$080B &    329.3896 & $    -7.7943$ & C & 0.0584 & 430 &  $92.43 \pm 1.16$ &   $13.88 \pm 0.71$ &   $6.47 \pm 0.92$ &   $5.67 \pm 0.96$ &   $ -3.42$ &  $ -0.78$ &  $ -0.20$ &  $ 0.00$ & $ 4.89$\\
2228$-$087 &    337.8701 & $    -8.4849$ & P & 0.0810 & 107 &  $10.89 \pm 0.67$ &   $10.48 \pm 0.31$ &   $6.39 \pm 0.72$ &   $6.30 \pm 0.62$ &   $ -0.07$ &  $ -0.51$ &  $ -0.02$ &  $ 8.89$ & $ 6.20$\\
2229$-$086 &    337.9302 & $    -8.4088$ & C & 0.0810 & 812 &  $64.79 \pm 2.52$ &   $33.38 \pm 1.13$ &   $20.42 \pm 0.85$ &   $16.84 \pm 0.70$ &   $ -1.19$ &  $ -0.50$ &  $ -0.29$ &  $ 1.98$ & $13.56$\\
2247+106B &    342.5818 & $    10.9034$ & $-1$ & 0.0768 & 14 &  $2.44 \pm 0.24$ &   $0.99 \pm 0.28$ &   $3.07 \pm 1.08$ &   $<1.11$ & $ -1.62$ &  $  1.16$ &  $< -1.55$ &  $ 0.02$ & \nodata\\
2321+164 &    350.9762 & $    16.6804$ & $-1$ & 0.0416 & 46 &  $20.40 \pm 0.57$ &   $9.95 \pm 0.44$ &   $6.96 \pm 0.10$ &   $2.30 \pm -1.00$ &   $ -1.29$ &  $ -0.37$ &  $ -1.69$ &  $ 0.47$ & $ 0.66$\\
2322+143A(1) &    351.1339 & $    14.6396$ & C & 0.0421 & 187 &  $65.59 \pm 1.36$ &   $24.47 \pm 0.80$ &   $6.75 \pm 0.92$ &   $4.94 \pm 0.78$ &   $ -1.78$ &  $ -1.32$ &  $ -0.48$ &  $ 0.37$ & $ 3.47$\\
2322+143B(1) &    351.1548 & $    14.6425$ & C & 0.0421 & 76 &  $25.85 \pm 0.95$ &   $14.01 \pm 0.54$ &   $6.84 \pm 1.05$ &   $3.15 \pm 0.71$ &   $ -1.10$ &  $ -0.74$ &  $ -1.18$ &  $ 1.03$ & $ 1.32$\\
2322$-$123 &    351.3324 & $   -12.1241$ & P & 0.0852 & 1699 &  $415.59 \pm 0.91$ &   $205.83 \pm 0.66$ &   $59.58 \pm 1.22$ &   $31.26 \pm 0.99$ &   $ -1.27$ &  $ -1.27$ &  $ -0.99$ &  $10.31$ & $15.13$\\
2332+270(2) &    353.7570 & $    27.3714$ & $-1$ & 0.0617 & 61 &  $9.23 \pm 0.45$ &   $<1.02$ & $<1.47$ & $<1.90$ & $< -3.97$ &  \nodata & \nodata & \nodata & \nodata\\
2333+208(1) &    354.1270 & $    21.1466$ & P & 0.0569 & 55 &  $11.41 \pm 0.33$ &   $6.60 \pm 0.27$ &   $4.30 \pm 0.62$ &   $4.25 \pm 0.74$ &   $ -0.99$ &  $ -0.44$ &  $ -0.02$ &  $ 0.64$ & $ 4.19$\\
2333+208(3) &    354.1654 & $    21.1021$ & P & 0.0569 & 55 &  $2.33 \pm 0.51$ &   \nodata & \nodata & \nodata & \nodata & \nodata & \nodata & \nodata & \nodata\\
2335+267(1) &    354.6225 & $    27.0314$ & C & 0.0321 & 7650 &  $267.48 \pm 15.60$ &   $212.63 \pm 5.08$ &   $139.12 \pm 2.16$ &   $102.95 \pm 1.33$ &   $ -0.41$ &  $ -0.44$ &  $ -0.46$ &  $79.97$ & $73.37$\\
2335+267(2) &    354.6330 & $    27.0244$ & $-1$ & 0.0321 & 7650 &  $316.87 \pm 22.30$ &   $184.23 \pm 12.60$ &   $<109.90$ & \nodata & $ -0.98$ &  $< -0.53$ &  \nodata & $18.27$ & \nodata\\
2348+058 &    357.7107 & $     6.1495$ & P & 0.0556 & 50 &  $8.49 \pm 0.29$ &   $3.02 \pm 0.24$ &   $5.28 \pm 1.88$ &   $<1.77$ & $ -1.86$ &  $  0.57$ &  $< -1.67$ &  $ 0.04$ & \nodata\\
2352+261(1) &    358.8519 & $    26.4047$ & C & 0.2404 & 311 &  $3.89 \pm 0.84$ &   $1.48 \pm 0.57$ &   $1.91 \pm 0.82$ &   $<3.25$ & $ -1.74$ &  $  0.26$ &  $<  0.81$ &  $ 0.02$ & \nodata\\
2352+261(3) &    358.8247 & $    26.4162$ & $-1$ & 0.2404 & 311 &  $28.44 \pm 1.58$ &   $12.30 \pm 0.85$ &   \nodata & \nodata & $ -1.51$ &  \nodata & \nodata & $ 0.35$ & \nodata
\enddata


\tablenotetext{a}{Source name from \citet{owen97}, which is based on B1950 positions; Parentheses denote multiple components.}
\tablenotetext{b}{Morphology of the sources: ``$-1$'' denotes extended source; ``C'' means core; ``P'' refers to point-like.}
\tablenotetext{c}{Redshift taken from \citet{owen97}.}
\tablenotetext{d}{1.4 GHz flux from \citet{owen97}.}
\tablenotetext{e}{Predicted flux at 90 GHz based on the fluxes at C \& X bands and $\alpha_{CX}$.}
\tablenotetext{f}{Predicted flux at 90 GHz based on the fluxes at K \& Q bands and $\alpha_{KQ}$.}

\label{tab:maindata}

\end{deluxetable*}

\clearpage
\end{landscape}

\end{document}